\documentclass{elsart}

\usepackage{graphics}
\usepackage{amsmath}
\usepackage{amssymb}
\newcommand{\ie}{{\it i.e.}}
\newcommand{\eg}{{\it e.g.}}
\newcommand{\ea}{{\it et al.}}
\newcommand{\gd}{\dot{\gamma}}
\newcommand{\mua}{\overline{\mu}_L}
\newcommand{\mub}{\overline{\mu}_H}
\newcommand{\E}{\mathcal{E}}
\newcommand{\vect}[1]{\underline{#1}}
\newcommand{\tensor}[1]{\underline{\underline{#1}}}
\newcommand{\deriv}[1]{\stackrel{\mbox{\scriptsize{$\triangle$}}}{#1}}
\newcommand{\grad}{\vect{\nabla}}
\newcommand{\Dxi}{d}

\begin{document}

\begin{frontmatter}

\title{Linear instability of planar shear banded flow of both diffusive and 
non-diffusive Johnson-Segalman fluids}

\author[UCL]{Helen J. Wilson\corauthref{cor}}
\ead{helen.wilson@ucl.ac.uk}
\author{\and}
\author[Man]{Suzanne M. Fielding}
\ead{suzanne.fielding@manchester.ac.uk}

\corauth[cor]{Corresponding author}
\address[UCL]{Department of Mathematics, University College London, 
Gower Street, London WC1E 6BT, UK}
\address[Man]{School of Mathematics, Lamb Building, The University of 
Manchester, Booth Street East, Manchester M13 9EP, UK}

\begin{abstract}
We consider the linear stability of shear banded planar Couette flow of the 
Johnson-Segalman fluid, with and without the addition of stress diffusion 
to regularise the equations. In particular, we investigate the
linear stability of an initially one-dimensional ``base'' flow, with a
flat interface between the bands, to two-dimensional perturbations
representing undulations along the interface. We demonstrate analytically 
that, for the linear stability problem, the limit in which diffusion tends 
to zero is mathematically equivalent to a pure (non-diffusive) 
Johnson-Segalman model with a material interface between the shear bands, 
provided the wavelength of perturbations being considered is long relative 
to the (short) diffusion lengthscale. 

For no diffusion, we find that the flow is unstable to long waves for almost 
all arrangements of the two shear bands. In particular, for any set of
fluid parameters and shear stress there is some arrangement of shear bands 
that shows this instability. Typically the stable arrangements of bands are 
those in which one of the two bands is very thin. Weak diffusion provides a 
small stabilising effect, rendering extremely long waves marginally stable. 
However, the basic long-wave instability mechanism is not affected by this,
and where there would be instability as wavenumber $k\to 0$ in the absence 
of diffusion, we observe instability for moderate to long waves even with 
diffusion. 

This paper is the first full analytical investigation into an instability first
documented in the numerical study of~\cite{Fie05}. Authors prior to that work 
have either happened to choose parameters where long waves are stable or used 
slightly different constitutive equations and Poiseuille flow, for which the 
parameters for instability appear to be much more restricted. 

We identify two driving terms that can cause instability: one, a jump
in $N_1$, as reported previously by Hinch \ea~\cite{Hin92}; and the second, a 
discontinuity in shear rate. The mechanism for instability from the second 
of these is not thoroughly understood. 

We discuss the relevance of this work to recent experimental observations of 
complex dynamics seen in shear-banded flows. 
\end{abstract}

\end{frontmatter}

\section{Introduction}

Complex fluids such as wormlike~\cite{BritCall97} and
onion~\cite{diat93} surfactants commonly show flow instabilities and
flow-induced transitions that lead to spatially heterogeneous, ``shear
banded'' states. In shear thinning wormlike micelles, for example,
homogeneous flow becomes unstable above a critical shear rate. The
system then separates into bands of differing viscosity and internal
structure, separated by an interface that has its normal in the
flow-gradient direction.  Widespread experimental observations of this
phenomenon have been made by flow birefringence~\cite{capp1995}, and
by NMR~\cite{Mai96} and ultrasound velocity
imaging~\cite{Becu.Manneville.ea04}. More recently, fluidity banding
has been reported in soft glassy
materials~\cite{RayMouBauBerGuiCou02}.  In bulk mechanical
measurements, the main signature of shear banding is a kink followed
by a plateau in the steady state flow curve.

Beyond this basic picture, an accumulating body of data reveals that
shear banded states can fluctuate. Early evidence came from unsteady
erratic~\cite{BanBasSoo00} or periodic~\cite{WFF98,WunColLenArnRou01}
fluctuations in the wall stress at an applied value of the shear
rate. More recent velocimetry experiments with enhanced spatial and
temporal resolution have unambiguously revealed fluctuations in the
interface between the
bands~\cite{Becu.Manneville.ea04,HolLopCal03,callnote,Salmon,SalmonB}. To
date, however, most theoretical studies have considered only a flat,
stationary interface. In this paper, therefore, we study analytically
the linear instability of shear banded flow with respect to small
undulations along the interface.

\begin{figure}
\begin{center}
\resizebox{80mm}{!}{\rotatebox{-90}{\includegraphics{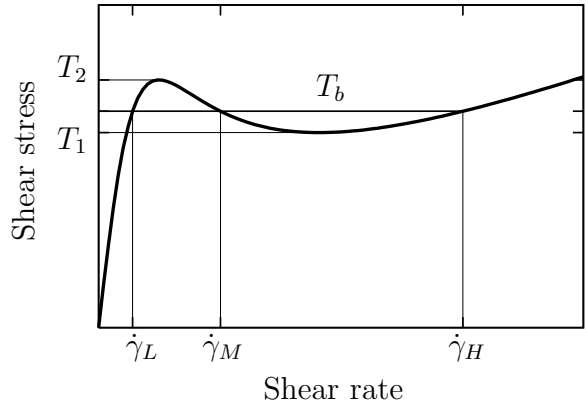}}}
\begin{picture}(0,0)
\put(-235,-100){\rotatebox{90}{Shear stress}}
\put(-218,-38){$T_2$}
\put(-218,-65){$T_1$}
\put(-120,-45){$T_b$}
\put(-140,-160){Shear rate}
\put(-192,-143){$\gd_L$}
\put(-163,-143){$\gd_M$}
\put(-70,-143){$\gd_H$}
\end{picture}
\end{center}
\caption{\label{fig:nonmonotone}A typical plot of shear stress against shear 
rate for steady, homogeneous shear flow of a shear-banding fluid. Here we use 
the Johnson-Segalman model with parameters $\epsilon=0.05$, $a=0.8$. If the 
shear stress is $T_b$, homogeneous shear flow with any of the shear rates 
$\gd_L<\gd_M<\gd_H$ is permitted; but flow with shear rate $\gd_M$ is unstable 
to one-dimensional perturbations.}
\end{figure}

Theoretically, shear banding is thought to arise from a
non-monotonicity in the underlying constitutive relation between the
shear stress and shear rate for homogeneous flow
\cite{Spe93,Mak95,Mai96,Mai97}. The simplest constitutive model to mimic this 
dependence (apart from ``toy'' models that do not obey the principle
of material frame indifference) is the Johnson-Segalman (JS)
model~\cite{Joh77}. A sample plot of shear stress against shear rate
in simple shear flow for this fluid is given in
figure~\ref{fig:nonmonotone}.  Homogeneous flow with a shear rate on
the decreasing part of the curve (e.g.~$\gd_M$) is unstable to
one-dimensional perturbations with wavevector in the flow-gradient
direction~\cite{Yer70}. The fluid therefore separates into a structure
comprising bands of differing shear rates $\gd_L$ and $\gd_H$, one on
each of the stable, upward-sloping parts of the curve.  The interface
between the bands has its normal in the flow-gradient direction. The
shear stress $T$ is uniform across the whole flow, as required by a
force balance.

The JS model in its original form contains no mechanism for uniquely
selecting the shear stress $T_b$ at which banding occurs. Instead, in a 
numerical study of shear banding in such a model, the shear stress in the 
steady banded state depends strongly on the startup 
history~\cite{Spe96,olmsted99a,malkus90,malkus91}, and can have a
stress anywhere in the range $T_1<T_b<T_2$ in
figure~\ref{fig:nonmonotone}. This conflicts notably with experiment,
which consistently reveals a highly reproducible banding stress.

It is therefore critical to regularise the model in some way, to
ensure stress selection. This is achieved by modifying the
constitutive equation to include a diffusive (``non-local'')
term~\cite{olmsted99a,lu99,olmstedlu97,olmsted92}. This mechanism was 
first proposed in 1989 by El-Kareh \& Leal~\cite{Kar89}, with the physical
interpretation that individual polymer molecules can slowly diffuse
across the interface, carrying their stress histories with them. Such
terms also arise naturally in models of liquid crystalline
dynamics. Regardless of their physical origin, non local terms lead to
the selection of a unique, reproducible shear banding stress, $T_b$,
as seen experimentally. They also provide a length scale for the
thickness of the interface. In contrast, in the local model the
interface is unphysically sharp: the flow variables jump
discontinuously across it.

The exact details of how stress diffusion should be added to the JS 
model vary from author to author. The most commonly used version 
is due to Olmsted \& coworkers~\cite{Rad00,Olm00}, in which a term 
proportional to the Laplacian of the extra stress, $\nabla^2\tensor{\Sigma}$, 
is added to the evolution equation for the polymeric stress. In this paper 
we will use a slightly more general form that is also capable of 
incorporating the model introduced by Yuan~\cite{Yua99a}, in which the 
term added is a negative multiple of the Laplacian of the rate of strain.

The theoretical framework just described has been developed largely in
the context of one dimensional (1D) studies that consider only the
flow gradient direction, normal to the interface between the
bands~\cite{Spe96,Esp96,Geo98}. Clearly, such studies assume from the
outset that the interface between the bands is perfectly flat and they
predict (with few exceptions:~\cite{Fie04,AraCat05}) time-independent
banded states. This is clearly at odds with the accumulating body of
data described above, revealing fluctuations of the banded state.

In view of this, a crucial question is whether the stationary, flat
banded state of 1D calculations will persist in 2D, or whether it
destabilises to exhibit large-amplitude interfacial fluctuations. The 
first step to answering this is clearly to perform a linear stability 
analysis of the 1D ``base state'' with respect to small 2D (flow, 
flow-gradient) perturbations corresponding to wavelike undulations along 
the interface.

This was first addressed within local models. McLeish~\cite{McL87}
considered a Doi-Edwards type fluid in capillary flow. He found
instability to long waves, provided the high shear rate band is very
narrow. As we shall see later, this is qualitatively very different
from our results. (McLeish did not give the specific parameters of his
calculation, so quantitative comparison is not possible.) 
Renardy~\cite{Ren95} examined the stability of the local JS model in
planar banded Couette flow.  She found linear instability for short
wavelengths (wavenumber greater than 8). For mainly historical
reasons, however, she happened to confine her study to a base state
corresponding to ``top-jumping'' ($T_b=T_2$) and an extremely thin 
high-shear band. We will return below to comment on this choice in the 
context of our own findings.

The first observation of linear instability within the diffusive JS
model was in the numerical study of~\cite{Fie05}.  This considered
general band thicknesses and demonstrated, for the first time,
instability with respect to long and moderate-length waves.  The
short-wave instability predicted in~\cite{Ren95} was eliminated by the
stabilising presence of diffusion. A subsequent non-linear numerical
study showed the interface to be restabilised at the level of finite
amplitude fluctuations~\cite{Fie06}.

The main contribution of the present study is a detailed analytical
interpretation of the numerical findings of~\cite{Fie05}.  We
start by deriving the important result that, for the limit of weak
diffusion, the 2D linear stability properties of the diffusive JS
model are equivalent to those of the original local model, with a
``material interface'' (defined below) between the bands. This
equivalence was not obvious {\it a priori}, since the 1D behaviour
differs so dramatically between the two models: as noted above, the
diffusionless limit is singular in 1D because it has no mechanism for
selecting the banding stress. The addition of weak stress diffusion
thus drastically modifies the 1D global properties, by selecting a
unique base state out of the continuum of possibilities.

This equivalence with the local model allows us to simplify
considerably the 2D stability analysis of the diffusive case. In
consequence, we are able to plot out the full spinodal boundary of
instability in the phase diagram, and to predict the dispersion
relations seen numerically. For a very wide range of model parameters, 
we find instability to waves of moderate wavelength $\lambda$ having 
$h^2 \ll \lambda^2 < L^3/h$ for small diffusion length, $h$ (where $L$ is 
a typical channel lengthscale). We also identify two possible driving terms 
for instability: one due to a jump across the interface in the base state 
shear rate, the other due to a jump in base state normal stress.

The paper is structured as follows. In section~\ref{sec:goveq} we
introduce our governing equations; in \S\ref{sec:steady} we lay out what is 
known about the one-dimensional steady state solution of these equations. In 
\S\ref{sec:linear} we set up the two-dimensional linear stability problem, and 
in \S\ref{sec:analysis-small-l} show analytically that, in the 
limit of small diffusion, for perturbations whose wavelength is not 
asymptotically short, the region between the shear bands may be considered as 
a material interface. This allows us to work, for the remainder of the paper,
with the simpler non-diffusive model (but with a value for the selected stress 
in the one-dimensional base state, $T_b$, as selected by 
diffusive terms). A long-wave stability analysis is given in 
\S\ref{sec:longwave}, and the results of a full numerical calculation in 
\S\ref{sec:num}. In \S\ref{sec:conc} we review our results and draw 
conclusions. 

\section{Governing equations and dimensionless form}
\label{sec:goveq}

The standard equations governing flow of an incompressible inertialess fluid 
are conservation of mass and a force balance:
\begin{equation}
\nabla\cdot\vect{U} = 0 \hspace{10mm} \nabla\cdot\tensor{S} = 0  
\end{equation}
in which $\vect{U}$ is the velocity field and $\tensor{S}$ is the total stress
tensor. The stress consists of an isotropic pressure $P$, a Newtonian solvent 
term of viscosity $\eta$ plus a polymer extra stress $\tensor{\Sigma}$: 
\begin{equation}
\tensor{S} = -P\tensor{I} + 2\eta\tensor{E} + \tensor{\Sigma}.
\end{equation}
The extra stress evolves with dynamics based on the Johnson-Segalman 
equation but with two possible diffusion terms added to regularise the 
equations: 
\begin{equation}
\deriv{\tensor{\Sigma}} = 2G\tensor{E} 
- \frac{1}{\tau}\tensor{\Sigma}  
- 2\Gamma \nabla^2\tensor{E}
+ {\mathcal D}\nabla^2\tensor{\Sigma}.
\end{equation}
The case ${\mathcal D}=0$ reproduces Yuan's model \cite{Yua99a}; the case 
$\Gamma=0$ reproduces the more usual model of Olmsted \& 
coworkers~\cite{Rad00,Fie05}; and if ${\mathcal D} = \Gamma=0$ we regain the 
original Johnson-Segalman model~\cite{Joh77}.

The derivative is the Johnson-Segalman form, with ``slip parameter'' $a$: 
\begin{equation}
\deriv{\Sigma_{ij}} = 
\left(\frac{\partial}{\partial t} + U_k\nabla_k\right)\Sigma_{ij}
- \left[\Sigma_{ik}\Omega_{kj} - \Omega_{ik}\Sigma_{kj} 
+ a(E_{ik}\Sigma_{kj} + \Sigma_{ik}E_{kj})\right],
\end{equation}
and the tensors $\tensor{E}$ and $\tensor{\Omega}$ are based on the velocity 
gradient: 
\begin{equation}
E_{ij} = \half(\nabla_i U_j + \nabla_j U_i) \hspace{10mm}
\Omega_{ij} = \half(\nabla_i U_j - \nabla_j U_i).
\end{equation}
We scale lengths with the channel width $L$, times with the polymer relaxation 
time $\tau$, and stresses with the modulus $G$. The dimensionless governing 
equations are then:
\begin{equation}
\nabla\cdot\vect{U} = 0 \hspace{10mm} \nabla\cdot\tensor{S} = 0.  
\end{equation}
\begin{equation}
\tensor{S} = -P\tensor{I} + 2\epsilon\tensor{E} + \tensor{\Sigma}
\end{equation}
\begin{equation}
\deriv{\tensor{\Sigma}} = 2\tensor{E} - \tensor{\Sigma}  
+ l^2\left\{
- 2\Delta\nabla^2\tensor{E} + (1-\Delta)\nabla^2\tensor{\Sigma}
\right\}.
\end{equation}
We have used three dimensionless parameters:
\begin{equation}
\epsilon = \eta/G\tau \hspace{10mm} 
l^2\Delta = \Gamma G/L^2 \hspace{10mm} 
l^2(1-\Delta) = {\mathcal D}\tau/L^2. 
\end{equation}
Thus $\epsilon$ is the retardation parameter, $l$ is a characteristic 
diffusion lengthscale, and $\Delta$ is a selection parameter to choose between 
the different stress-diffusion mechanisms. The dimensional diffusion length 
$h$ is $lL$. Shear-banding is possible for $\epsilon < 0.125$ and in all our 
examples we will use the value $0.05$; it is more difficult to predict the 
likely physical values of other parameters, so we consider ranges 
$0\le l \le0.01$, $0.3 \le a \le 0.8$ and $0 \le \Delta \le 1$. 

We will be considering a planar shear flow, bounded between two solid walls. 
As well as the standard velocity boundary conditions of no slip and no 
penetration at the walls, we require boundary conditions on the stress 
because of the higher derivatives introduced by the diffusion terms. We have 
chosen to use a boundary condition imposing no flux of extra stress at the 
boundaries: for a wall given by $y=\mathrm{constant}$ this imposes the 
condition $\partial\tensor{\Sigma}/\partial y = \tensor{0}$ at the wall. 
Physically this constrains our steady flow not to have boundary layers near 
the walls, although (as in \S\ref{sec:inner}) boundary layer structures are 
permitted within the body of the fluid.  

\section{Steady-state solution}
\label{sec:steady}

If we impose the constraint of a steady unidirectional flow with no pressure 
gradient along the channel, then all flow variables depend only on the position
across the channel, $y$. We denote differentiation with respect to $y$ by $D$. 
We also introduce new variables $T_{ij}$ based on the stress tensor 
$\Sigma_{ij}$: 
\begin{eqnarray}
T_{11} &=& \half(1-a) \Sigma_{11} - \half(1+a) \Sigma_{22} \label{eq:baseT11}\\
T_{22} &=& \half(1-a) \Sigma_{11} + \half(1+a) \Sigma_{22} \\
T_{12} &=& \Sigma_{12}.
\end{eqnarray}
The equations governing the flow then reduce to 
\begin{equation}
U_x = U(y) \hspace{10mm} U_y = 0 \hspace{10mm} \gd = DU
\end{equation}
for the velocity field, and 
\begin{eqnarray}
S_{11} &=& -P + (1-a)^{-1}(T_{22}+T_{11}) \\ 
S_{12} &=& \epsilon\gd + T_{12} \\
S_{22} &=& -P + (1+a)^{-1}(T_{22}-T_{11}) 
\end{eqnarray}
for the total stress. Finally, for the polymer stress we have 
\begin{equation}
l^2(1-\Delta)D^2T_{22} - T_{22} = 0
\end{equation}
which is satisfied if $T_{22}=0$, and 
\begin{eqnarray}
(1-a^2)\gd T_{12} &=& - l^2(1-\Delta) D^2T_{11} + T_{11} \\
\gd T_{11} &=& \gd - T_{12} - l^2\Delta D^2\gd + l^2(1-\Delta) D^2T_{12}.
\label{eq:basegdT11}
\end{eqnarray}
In Couette flow, there is no pressure gradient along the channel:
\begin{equation}
\frac{\partial}{\partial x}P = 0,
\end{equation}
and the unscaled momentum equations are 
\begin{equation}
T_{12}+\epsilon\gd = T_b
\end{equation}
where we recall that the shear stress $T_b$ is selected from the continuum of
possibilities $T_1\le T_b\le T_2$ only in the diffusive model~\cite{Rad00}, 
and
\begin{equation}
\frac{\partial P}{\partial y} = -(1+a)^{-1}DT_{11}.
\end{equation}
This last equation is the only place that the dependence of $P$ on $y$ 
appears, so we can solve it by setting
\begin{equation}
P = -(1+a)^{-1}T_{11}.
\end{equation}

Now let us look at two cases: $l$ small but non-zero (diffusive JS), and 
$l=0$ (non-diffusive). If $l=0$ we have the Johnson-Segalman model in its 
original form, with no mechanism to select the shear stress: in this 
(unphysical) case, shear banding can occur at any shear stress in the 
interval $T_1<T_b<T_2$ (see figure~\ref{fig:nonmonotone}). For any such 
value, there are three possible shear rates 
$\gd_L(T_b)<\gd_M(T_b)<\gd_H(T_b)$, of which a flow with $\gd_M$ would be 
unstable to one-dimensional perturbations. The system will therefore consist 
of bands of the two shear rates $\gd_L$ and $\gd_H$. There may be several 
of these.

If we allow $l\not=0$, the system changes radically. The shear stress 
$T_b=T_\mathrm{sel}$ is uniquely selected~\cite{Yua99,Rad00}, and the flow 
will separate into shear bands of $\gd_L$ and of $\gd_H$. Between these bands, 
rather than a sharp interface, is a matching region of width $l$, across which 
the flow variables vary continuously.

In order to carry out direct comparisons between the $l=0$ and $l\not=0$ 
cases, we will assume here that only one region of each shear rate is formed, 
and that there is then only one interface between high- and low-shear rate 
bands. 

For definiteness, we assume that there is a band of low shear rate $\gd_L$ 
near the wall $y=0$, which continues up to a position $y=\kappa$. Near the 
wall $y=1$ (and continuing down towards $y=\kappa$) there is a band of high 
shear rate $\gd_H$. The average shear rate (and hence the speed of the upper 
wall) is $U_\mathrm{wall} = \kappa\gd_L + (1-\kappa)\gd_H$. This can equally 
be considered as the Weissenberg number in our nondimensionalisation. The 
symmetry of planar Couette flow is such that these two bands could be 
interchanged to produce an essentially equivalent flow: we restrict our 
attention to the arrangement with the low-shear band near $y=0$. 

For $l\not=0$, the two regions well away from the matching region will be 
denoted the {\em outer}, and the matching region of width $l$, the {\em inner} 
region. Each outer region is undergoing homogeneous shear flow, while the 
inner region can be thought of as a finite-width interface between these two 
phases. 

\subsection{Outer solution}

Well away from the matching region, we expect derivatives of all our 
quantities to be at most order 1: so if $l$ is very small we can neglect 
terms $l^2D^2$ and regain the equations governing one-dimensional flow of 
the original Johnson-Segalman equation. This system has been investigated 
thoroughly \cite{Rad00,Geo98}. The stress components of the solution are 
\begin{equation}
T_{11} = \frac{(1-a^2)\gd^2}{1+(1-a^2)\gd^2} \hspace{10mm}
T_{12} = \frac{\gd}{1+(1-a^2)\gd^2} \hspace{10mm}
T_{22} = 0
\end{equation}
and the shear stress condition $T_{12}=T_b-\epsilon\gd$ gives a cubic equation 
for $\gd$: 
\begin{equation}
(1-a^2)\epsilon\gd^3 - (1-a^2)T_b\gd^2 + (1+\epsilon)\gd - T_b = 0. 
\label{eq:cubic}
\end{equation}
Since the neglected terms all involve derivatives of the leading-order terms, 
which are constants, this is an exact solution to the governing equations for 
each permissible value of $\gd$. 

If $a^2<1$ and $\epsilon<1/8$, there is a range of values of the shear stress 
$T_b$ for which $T_1<T_b<T_2$ and equation (\ref{eq:cubic}) has three 
solutions $\gd_L$, $\gd_M$ and $\gd_H$, which allows the possibility of shear 
bands. 

\subsection{Inner region}
\label{sec:inner}

If $l$ is small but non-zero, we do not expect a truly sharp interface 
between regions of high and low shear rates. Instead, in the inner region, 
the second-order derivatives become $O(l^{-2})$ and so become large enough 
to be important, despite being multiplied by $l^2$. 

We take this region to be a layer centred on $\kappa$ and use a rescaling 
variable $y = \kappa + l\xi$. The equations become (denoting derivative 
with respect to $\xi$ by $\Dxi$):
\begin{eqnarray}
(1-a^2)\gd T_{12} &=& -(1-\Delta) \Dxi^2T_{11} + T_{11} \\
\gd T_{11} &=& \gd - T_{12} - \Delta \Dxi^2\gd + (1-\Delta) \Dxi^2T_{12} \\
T_{12}+\epsilon\gd &=& T_b \label{eq:shear}
\end{eqnarray}
with matching conditions at the edges of the layer for $\gd$, $T_{11}$ and 
$T_{12}$ in terms of the outer solution. 
Existence of a solution to this system which matches onto the two 
constant-$\gd$ solution branches (with $\gd$ dependent on $T_b$) as 
$\xi\to\pm\infty$ is a constraint on $T_b$. 

This is most easily seen in the case $\Delta=1$, in which, following Lu, 
Olmsted \& Ball~\cite{lu99}, (\ref{eq:shear}) can be written as 
\begin{equation}
\Dxi^2\gd = \gd + (\epsilon\gd - T_b)(1 + (1-a^2)\gd^2).
\end{equation}
Multiplying by $\Dxi\gd$ and integrating with respect to $\xi$ across the 
interface region gives a continuity condition 
\begin{equation}
\left[\frac{1}{2}(1+\epsilon)\gd^2 + \frac{1}{4}\epsilon(1-a^2)\gd^4 - T_b\left(\gd + \frac{1}{3}(1-a^2)\gd^3\right) \right]_{\gd_L}^{\gd_H} = 0
\end{equation}
which, coupled with the fact that $\gd_L$ and $\gd_H$ depend on $T_b$, has a 
unique physically relevant solution 
\begin{equation}
T_b(\epsilon,a) = \frac{3}{2}(1-a^2)^{-1/2}(2\epsilon - 4\epsilon^2)^{1/2}. 
\label{eq:yuanTb}
\end{equation} 
A similar result (with a different function $T_b(\epsilon,a)$) is true for 
other values of $\Delta$. For reference, we note that at $\epsilon=0.05$, 
the extremal selected values are
\begin{equation}
\Delta=0:\hspace{2mm} T_b = 0.48284(1-a^2)^{-1/2}; \hspace{10mm} 
\Delta=1:\hspace{2mm} T_b = 0.45(1-a^2)^{-1/2}.
\label{eq:Tbsel}
\end{equation}
In this way, the addition of weak diffusion is seen to regularise the 
equations by selecting one possible value $T_b$ out of the continuum of 
possibilities $T_1 < T_b < T_2$ at which a one-dimensional shear-banded 
solution can occur. This one-dimensional solution will be taken as the relevant
base state about which we analyse two-dimensional perturbations (in the 
flow, flow gradient plane) in the remainder of the paper. 

\section{Linearised equations}
\label{sec:linear}

We now add a small perturbation proportional to $\E$, such that 
\begin{equation}
\vect{U} = \left(\begin{array}{cc} U + u\E, & v\E \end{array}\right) 
\hspace{10mm}
P = P_\mathrm{base} + p\E, 
\end{equation}
\begin{equation}
\tensor{S} = \left(\begin{array}{cc} S_{11}+s_{11}\E & S_{12}+s_{12}\E \\ 
S_{12} + s_{12}\E & S_{22} + s_{22}\E\end{array}\right) 
\hspace{10mm}
\tensor{T} = \left(\begin{array}{cc} 
T_{11}+t_{11}\E & T_{12}+t_{12}\E \\ 
T_{12} + t_{12}\E & t_{22}\E\end{array}\right) 
\end{equation}
in which
\begin{equation}
\E = \delta\exp{(ikx - i\omega t)}
\end{equation}
for a small parameter $\delta$, and we will use $D$ to denote differentiation 
with respect to $y$. We have introduced the dimensionless wavenumber 
$k=2\pi L/\lambda$ for waves of wavelength $\lambda$. We can now linearise 
the equations about the base state in this small parameter, to obtain our new 
governing equations for evolution of the linear system. For a fixed real 
wavenumber $k$, if we obtain a valid solution with $\mathrm{Im}(\omega)>0$ 
then the system is linearly unstable. This analysis is necessarily restricted 
to considering instabilities which manifest at linear order; any instability 
(like the high-Reynolds number instability of laminar flows) which appears 
only at nonlinear order will not be captured here. The mass conservation 
equation is 
\begin{equation}
iku + Dv = 0. 
\end{equation}
This is automatically satisfied by the introduction of a streamfunction $\psi$ 
with  $v=-ik\psi$ and $u=D\psi$, such that the perturbation velocity 
$\vect{u}\E=\grad\times(\psi\vect{\hat{z}}\E)$. The remaining governing 
equations for the force balance become 
\begin{equation}
iks_{11} + Ds_{12} = 0 \hspace{10mm}
iks_{12} + Ds_{22} = 0 \label{eq:goveq_first} 
\end{equation}
\begin{eqnarray}
s_{11} &=& -p + 2\epsilon ikD\psi + (1-a)^{-1}(t_{22}+t_{11})\label{eq:s11}\\
s_{12} &=& \epsilon(D^2\psi + k^2\psi) + t_{12} \label{eq:s12}\\
s_{22} &=& -p - 2\epsilon ikD\psi + (1+a)^{-1}(t_{22}-t_{11})\label{eq:s22}
\end{eqnarray}
and for the polymer stress we have:
\begin{multline}
(-i\omega + ikU + 1 + l^2 k^2)t_{22} = 
l^2 (1-\Delta)D^2 t_{22} + 2ikal^2\Delta(D^2-k^2)D\psi \\ \mbox{}
+ 2k^2a\psi T_{12} + 2ika (T_{11}-1) D\psi \label{eq:t22}
\end{multline}
\begin{multline}
(-i\omega + ikU + 1 + l^2 k^2)t_{11} = 
l^2(1-\Delta)D^2t_{11} - 2ikl^2\Delta(D^2-k^2)D\psi \\ \mbox{}
+ (1-a^2)\gd t_{12} 
+ ik\psi DT_{11} 
+ 2ikD\psi 
+ T_{12}[(1-a^2)D^2\psi - k^2(1+a^2)\psi] \label{eq:t11}
\end{multline}
\begin{multline}
(-i\omega + ikU + 1 + l^2 k^2)t_{12} = 
l^2(1-\Delta)D^2t_{12} - l^2\Delta(D^4-k^4)\psi \\ \mbox{}
- \gd t_{11} 
+ ik\psi DT_{12}
+ (1-T_{11})(D^2+k^2)\psi
+ 2(1-a^2)^{-1}k^2\psi T_{11}. \label{eq:goveq_last}
\end{multline}

\section{Equivalence of the stability analyses at $l\to 0$ and $l=0$}
\label{sec:analysis-small-l}

\subsection{Introduction}

The limit $l\to 0$ is a singular limit, in the sense that the coefficient of 
the highest derivative in the governing equations is multiplied by $l^2$. 
As discussed above, therefore, the one-dimensional solutions to equations 
(\ref{eq:baseT11}--\ref{eq:basegdT11}) are very different for small $l$ from 
$l=0$: for $l\not=0$ the shear stress $T_b$ at which a banded state can exist 
is uniquely selected, whereas for $l=0$ banding can occur at any shear stress 
$T_1<T_b<T_2$. 

In this section we consider the limit of very small $l$ in the two-dimensional 
stability problem, and show that, in contrast to the one-dimensional case, the 
limit $l\to 0$ is equivalent to $l=0$ (with suitable interfacial boundary 
conditions), for any given base state. To do this, we start by considering the 
full diffusive model, in the asymptotic limit of small $l$. As $l\to 0$ the 
diffusion terms become unimportant everywhere in the flow apart from in the 
inner region, which has dimension $l$. This region separates two phases of 
flow at different shear rates, in which the governing equations 
are those of the original (non-diffusive) Johnson-Segalman equation. 

The crucial question to address here is what boundary conditions are imposed 
at the edge of each of these phases as a result of the structure of the inner 
solution. Our central finding is that these conditions are the same in the 
limit of a thin interface as they would be for the case of an infinitesimally 
sharp material interface (that is, an interface across which material does not 
pass) with $l=0$. Accordingly, we now summarise the case of 
a sharp (diffusion-free) material interface $l=0$ before proceeding with our 
analysis of weak diffusion, $l\to 0$. 

McLeish~\cite{McL87} has argued that if the functional relating current stress 
to strain history is well-behaved, a small perturbation to the flow can only 
cause a small perturbation to the stress in any fluid element. Thus material 
can never be transported across a sharp interface since this would cause an 
order 1 change in the stress for that material. 

This physical argument leads to a kinematic boundary condition: if the 
interface is defined by points $y = \overline{\eta} + \zeta(x,\overline{\eta},t)$ (where $\overline{\eta}$ is
the Lagrangian cross-channel coordinate) then 
\begin{equation}
  \frac{D\overline{\eta}}{Dt} = \frac{\partial\overline{\eta}}{\partial t} 
  + \vect{u}\cdot\vect{\nabla}\overline{\eta} = 0.
\end{equation}
This boundary condition is standard for an interface separating two different 
materials: in particular, Renardy~\cite{Ren95} applied it to two phases of 
a shear-banded Johnson-Segalman fluid. For our linear perturbation problem,
if the interface $\overline{\eta}=\kappa$ is displaced from $y=\kappa$ to 
$y=\overline{\eta}+\zeta\E$, then since $\overline{\eta}=y-\zeta\E$, the kinematic boundary 
condition (correct to linear order) is expressed as 
\begin{equation}
(-i\omega + ikU(\kappa))\zeta = -ik\psi(\kappa).
\end{equation}
Then imposing continuity of velocity and traction across the interface gives 
boundary conditions for the jumps in $\psi$, $D\psi$ and $s_{ij}$ across the 
interface.

With this summary of the diffusion-free sharp material interface in mind, we 
now proceed to show that the case $l\to 0$ is equivalent to it.

\subsection{Asymptotic analysis}
\label{sec:lowl}

We now consider the asymptotic limit $l\to 0$. We scale our perturbation flow 
so that the values of standard quantities (\eg\ $\psi$, $D\psi$, $t_{12}$) are 
order 1 in the outer regions. Within the inner region we pose the following 
asymptotic series (in which superscripts on $\psi$, $s_{ij}$, $t_{ij}$ and $p$ 
are for labelling only, while those on $l$ indicate powers):
\begin{equation}
\psi \sim \psi^0 + l\psi^1 + l^2\psi^2 + \cdots \hspace{10mm}
t_{ij} \sim l^{-1}t_{ij}^{-1} + t_{ij}^0 + \cdots 
\end{equation}
\begin{equation}
p \sim l^{-1}p^{-1} + p^0 + \cdots  \hspace{10mm}
s_{ij} \sim l^{-1}s_{ij}^{-1} + s_{ij}^0 + \cdots  
\end{equation}
These scalings are not obvious {\it a priori}; rather, the existence of a 
solution with these scalings will justify its choice. 
As before, we scale lengths within the interfacial region as 
$\xi = (y-\kappa)/l$, giving $lD = \Dxi$. This analysis will be valid as 
long as this lengthscale is well separated from other lengthscales in the 
flow: in particular, the arguments and analysis below will not apply for 
very short waves for which $k\sim l^{-1}$: we restrict our analysis to 
the case $kl \ll 1$. 

We also pose a series for the eigenvalue $\omega$:
\begin{equation}
\omega \sim \omega_0 + l\omega_1 + \cdots   
\end{equation}

Well away from the interface, all quantities are at most order 1, which we 
have ensured by scaling the whole perturbation. This gives a series of 
conditions on our variables as $\xi\to\pm\infty$. In particular, since 
$D\psi$, $t_{ij}$ and $s_{ij}$ must be finite, with no contribution at 
$O(l^{-1})$, we have the conditions 
\begin{equation}
\Dxi\psi^0 \to 0, \mbox{ } t_{ij}^{-1} \to 0 
 \mbox{ and } s_{ij}^{-1} \to 0 \mbox{ as } \xi \to \pm \infty.
\label{eq:match}
\end{equation}
We split the governing equations into successive orders of $l$. At order 
$l^{-2}$ we have, from the stress equations, simply (from (\ref{eq:s12}))
\begin{equation}
\Dxi^2\psi^0 = 0   \label{eq:psi0}
\end{equation}
which, along with the matching condition (\ref{eq:match}) gives 
$\psi^0 = \mbox{constant} = \alpha_0$. This means that the streamfunction 
$\psi$ (or velocity in the $y$-direction) is continuous across the interface 
region, and $\alpha_0$ is the value of $\psi$ at the interface.

At $O(l^{-2})$ in the momentum equations, we have: 
\begin{equation}
\Dxi s_{12}^{-1} = 0    \hspace{10mm} \Dxi s_{22}^{-1} = 0      
\end{equation}
which, along with (\ref{eq:match}), give $s_{12}^{-1}=s_{22}^{-1}=0$. Then 
the stress equations (\ref{eq:s11})--(\ref{eq:s22}) at $O(l^{-1})$ give 
\begin{equation}
p^{-1} = -(1+a)^{-1}t_{11}^{-1}, \hspace{10mm} 
s_{11}^{-1} = 2(1-a^2)^{-1}t_{11}^{-1}, 
\label{eq:lzeros11}
\end{equation}
\begin{equation}
\epsilon \Dxi^2\psi^1 + t_{12}^{-1} = 0. 
\label{eq:ltozero1}
\end{equation}
The equations governing $t_{ij}^{-1}$ at order $O(l^{-1})$ become: 
\begin{multline}
(-i\omega_0 + ikU(\kappa) + 1)t_{11}^{-1} = 
(1-\Delta)\Dxi^2t_{11}^{-1} 
+ (1-a^2)\gd t_{12}^{-1} \\
+ ik\alpha_0 \Dxi T_{11} 
+ (1-a^2)T_{12} \Dxi^2\psi^1 
\end{multline}
\begin{multline}
(-i\omega_0 + ikU(\kappa) + 1)t_{12}^{-1} = 
(1-\Delta)\Dxi^2t_{12}^{-1} - \Delta \Dxi^4\psi^1
- \gd t_{11}^{-1} \\ \mbox{}
+ ik\alpha_0 \Dxi T_{12}
+ (1-T_{11})\Dxi^2\psi^1 
\label{eq:ltozero3}
\end{multline}
\begin{equation}
t_{22}^{-1}=0.
\end{equation}
We now introduce a quantity 
\begin{equation}
\zeta = -ik\alpha_0/(-i\omega_0 + ikU(\kappa)).  
\end{equation}
Equations (\ref{eq:ltozero1})--(\ref{eq:ltozero3}) then give the following 
system: 
\begin{equation}
\epsilon \Dxi^2\psi^1 + t_{12}^{-1} = 0 
\end{equation}
\begin{multline}
t_{11}^{-1} = (1-\Delta)\Dxi^2t_{11}^{-1} + (1-a^2)\gd t_{12}^{-1}
+ (1-a^2)T_{12}\Dxi^2\psi^1 \\
- (-i\omega_0 + ikU(\kappa))[t_{11}^{-1} + \zeta \Dxi T_{11}] 
\end{multline}
\begin{multline}
t_{12}^{-1} = (1-\Delta)\Dxi^2t_{12}^{-1} - \Delta \Dxi^4\psi^1
- \gd t_{11}^{-1} + (1-T_{11})\Dxi^2\psi^1
\\ \mbox{}
- (-i\omega_0 + ikU(\kappa))[t_{12}^{-1} + \zeta \Dxi T_{12}] 
\end{multline}
which is solved by 
\begin{equation}
\Dxi^2\psi^1 = -\zeta \Dxi\gd, \hspace{10mm}
t_{12}^{-1} = -\zeta \Dxi T_{12}, \hspace{10mm}
t_{11}^{-1} = -\zeta \Dxi T_{11}, \label{eq:dpsi1}
\end{equation}
all of which tend to zero for large $|\xi|$ as required. These are the first 
derivatives (in the flow-gradient direction) of the corresponding base-state 
quantities, which for long waves are the perturbations we would expect from a 
simple displacement of the interface by an amount $\zeta$. Integrating 
$\Dxi^2\psi^1$ gives 
\begin{equation}
\Dxi\psi^1 = \alpha_1 -\zeta \gd.  
\end{equation}
We note that 
(\ref{eq:lzeros11}) also gives 
\begin{equation}
s_{11}^{-1} 
= 2(1-a^2)^{-1}t_{11}^{-1}   
= -2(1-a^2)^{-1}\zeta \Dxi T_{11}
= -\zeta \Dxi S_{11}.
\end{equation}
At $O(l^{-1})$, the two momentum equations give 
\begin{equation}
iks_{11}^{-1} + \Dxi s_{12}^0 = 0  \hspace{10mm} 
iks_{12}^{-1} + \Dxi s_{22}^0 = 0  
\end{equation}
and thus
\begin{equation}
s_{12}^0 = \alpha_2 + ik\zeta S_{11}  \hspace{10mm} 
s_{22}^0 = \alpha_3. \label{eq:sij0}
\end{equation}

\subsection{Physical meaning}

We have now calculated all the singular components of the perturbation flow 
for the asymptotic limit $l\to 0$. What conclusions can we draw about the 
behaviour of the outer quantities (which are not directly affected by 
diffusion) at the edges of the inner layer? Looking at velocities and stresses 
(and recalling that $D\psi \sim l^{-1}\Dxi\psi$ so that the order 1 term in 
$D\psi$ comes from $\psi^1$), we have the following conditions (to within 
corrections of $O(l)$): 
\begin{equation}
\psi(\kappa_-) = \psi(\kappa_+) = \alpha_0  
\label{eq:psi_cont}
\end{equation}
which follows from (\ref{eq:psi0}), 
\begin{equation}
D\psi(\kappa_-) + \zeta\gd(\kappa_-) 
= D\psi(\kappa_+) + \zeta\gd(\kappa_+) = \alpha_1 
\label{eq:Dpsi_jump}
\end{equation}
from (\ref{eq:dpsi1}), and two conditions from (\ref{eq:sij0}): 
\begin{equation}
s_{12}(\kappa_-) - ik\zeta S_{11}(\kappa_-)
= s_{12}(\kappa_+) - ik\zeta S_{11}(\kappa_+) = \alpha_2 
\label{eq:s12_jump}
\end{equation}
\begin{equation}
s_{22}(\kappa_-) = s_{22}(\kappa_+) = \alpha_3
\label{eq:s22_cont}
\end{equation}
along with the definition 
\begin{equation}
\zeta = -ik\alpha_0/(-i\omega_0 + ikU(\kappa)) \hspace{10mm} 
(-i\omega_0 + ikU(\kappa))\zeta = -ik\psi(\kappa).
\label{eq:zeta_def}
\end{equation}
We have used the notation $X(\kappa_-)$ to denote the value of variable $X$ in 
the limit $y\to\kappa$ of the outer region given by $0\le y<\kappa$;  
similarly, $X(\kappa_+)$ denotes the value as $y\to\kappa$ in the outer region 
$\kappa<y\le 1$. 

These equations exactly reproduce those for a sharp material interface with 
displacement $\zeta$ from its original position and a discontinuity in both 
$\gd$ and $S_{11}$ across it in the base state. If we identify the quantity 
$\zeta$ as a displacement in the interface, then the conditions 
(\ref{eq:psi_cont}--\ref{eq:Dpsi_jump}) correspond to continuity of the total 
fluid velocity at the interface, and (\ref{eq:s12_jump}--\ref{eq:s22_cont}) 
come from the momentum equations, and correspond to continuity of the total 
traction $(\tensor{\Sigma} + \tensor{\sigma}\E)\cdot\vect{n}$ at the interface,
where $\vect{n}$ is the normal to the displaced interface. Of course the 
interface itself has width of order $l$, so only in the asymptotic limit 
$l\to 0$ do we truly have an interface to consider, which is why the analysis 
above is only valid in this limit. Note that the original Johnson-Segalman 
equations governing a linear perturbation are fourth-order in $\psi$, and 
equations (\ref{eq:psi_cont}--\ref{eq:Dpsi_jump}) supply four boundary 
conditions at the interface, as we require.

In this limit $l=0$ but with a finite interface displacement, we have 
effectively exceeded the validity of our original linearisation. When we 
linearised about the base state, we implicitly assumed that the scale of all 
perturbation velocities, displacements and so on was much smaller than any 
other quantity in the problem, and in particular that displacements must 
be smaller than the width of the interface region, $l$. In taking the 
asymptotic limit $l\to 0$ we have continued the assumption that displacements 
are small relative to the interface size: so this asymptotic limit is not 
founded on a well-posed physical problem. This makes it even more surprising 
that the limit should correspond exactly to the physical problem with an  
infinitely sharp material interface, in which the (small) displacement is 
much larger than the (zero) interface width. 

If we consider a perturbation to consist of an initial perturbation of the 
interface (this assumption is not always valid, as demonstrated recently by 
Kupferman~\cite{Kup05}), then the driving terms for the perturbation are 
the $\zeta$ terms in (\ref{eq:Dpsi_jump}) and (\ref{eq:s12_jump}), which 
depend on discontinuities in the base-state. If the shear rate were 
continuous across the interface, the only driving term would be the 
discontinuity in $S_{11}$ in (\ref{eq:s12_jump}); this would then allow us 
to recapture the normal stress instability mechanism elucidated by 
Hinch \ea~\cite{Hin92}. The discontinuity in shear rate in (\ref{eq:Dpsi_jump})
provides another, more complex, mechanism for instability, which is not 
yet fully understood. 

At order $l$, however, equations (\ref{eq:psi_cont}--\ref{eq:s22_cont}) will 
not be satisfied so the diffusion does alter the zero-transport property. This 
is to be expected if the stress diffusion derives physically from diffusive 
transport of individual polymer molecules carrying their stress history with 
them: we would expect material to travel across the interface, a distance of 
$l$, with diffusive velocity of $l^2$, on a timescale of order $l^{-1}$. This 
timescale corresponds to a contribution of order $l$ to $\omega$, \ie\ the 
next term in the asymptotic series, $\omega_1$. 

\subsection{Wall boundary conditions}

In our discussion of the boundary conditions applying across the diffusive 
interface layer, we have implicitly assumed that this layer is the only place 
in the perturbation flow where diffusion is important. However, the stress 
boundary conditions $D\sigma_{ij}=0$ at the walls are not automatically 
satisfied by the perturbation flow of a pure Johnson-Segalman fluid. We would 
expect that a stress boundary layer would form at each wall, matching between 
our outer solution and the no-flux boundary conditions. 

This does indeed occur, but the influence of these stress boundary layer 
regions does not extend into the bulk of the flow, and they do not affect 
the stability of the flow. They are standard boundary layers as one would 
expect for what is essentially an advection-diffusion equation in extra stress.
Although we are not presenting an analytical justification of the assertion 
that their effect is negligible, the close match between our results 
(neglecting the wall boundary layers) and those of \cite{Fie05} and of 
\S\ref{sec:sigma0} at small finite $l$ with no such assumptions, justifies 
our claim.

\section{Long wave analysis $k \ll 1$ at $l=0$} 
\label{sec:longwave}

We showed in Section~\ref{sec:analysis-small-l} that as long as the wavenumber 
$k$ is not very large (we assumed in Section~\ref{sec:lowl} that $kl\ll 1$), 
the interface between the two phases of the flow may be treated as a material 
interface, with corrections appearing at $O(l)$ for small $l$. A typical 
estimate of $l$ (the ratio of mesh size to channel width in~\cite{Fie03a}) is 
$2\times 10^{-4}$, so the waves which are excluded from our analysis are very 
short indeed. 

In summary, for small diffusion, the diffusive terms in the outer region do 
not affect the flow at any physically important level, so the only r\^ole 
played by diffusion is in the ``boundary'' conditions that it imposes on the 
outer at the edges of the inner, interface, region. We have just shown that 
these are equivalent to treating the interface as a material interface in the 
case $l=0$: so the solution to the $l=0$ problem with a material interface 
must be the same as the $l\to 0$ limit of the diffusive problem. This 
equivalence is verified in Section~\ref{sec:num} by the agreement between the 
$l=0$ results of this paper with earlier numerical results by one of the 
authors~\cite{Fie05} for small $l$. 

This finding allows us now to consider the simpler (although artificial) 
limit $l=0$, in which the diffusive terms play no part at all, and the base 
state quantities $\gd$, $T_{11}$ and $T_{12}$ are constant within each fluid 
phase, and treat the interface as a material surface. We are now dealing with 
the original Johnson-Segalman fluid~\cite{Joh77} and considering a stability 
problem which was studied by Renardy~\cite{Ren95} in 1995, although we have 
some analytical results for long waves where her work was purely numerical. 
However, as discussed above, for planar Couette flow of a Johnson-Segalman 
fluid in the non-monotonic region of the 
flow curve, there is no mechanism for selection of the shear stress. Any shear 
stress that intersects both the low- and high-shear-rate parts of the flow 
curve, $T_b$ with $T_1 < T_b < T_2$ (figure~\ref{fig:nonmonotone}) is a 
possible solution to the governing equations, with the choice of shear stress 
(for a given wall velocity) governing the position of the interface. Renardy's 
solution to this difficulty was to assume {\it top-jumping}, that is, the 
maximum value of the shear stress for which the flow curve is multi-valued. 
We, however, are using this local model as an approximation to the non-local 
case $l\not=0$, so we adopt the unique value of $T_b$ selected by use of the 
gradient terms in the stress-diffusive non-local case.  

In the case of long waves, $k\to 0$, $p$, $s_{11}$ and $s_{22}$ are larger than
the other perturbation quantities by a factor of $k^{-1}$, and the eigenvalue 
$\omega$ is expected to be of order $k$, based on prior experience of 
long-wave interfacial instabilities~\cite{Hin92}. If we scale these 
quantities accordingly: $p = k^{-1}\overline{p}$, 
$s_{ii}=k^{-1}\overline{s}_{ii}$, $\omega = k\overline{\omega}$, then the 
governing equations within each fluid, from 
(\ref{eq:goveq_first}--\ref{eq:goveq_last}), become, for the force balance:
\begin{equation}
i\overline{s}_{11} + Ds_{12} = 0 \hspace{10mm}
ik^2s_{12} + D\overline{s}_{22} = 0, 
\end{equation}
for the total stress:
\begin{eqnarray}
\overline{s}_{11} &=& -\overline{p} + 2\epsilon ik^2D\psi 
                                + (1-a)^{-1}k(t_{22}+t_{11}) \\
s_{12} &=& \epsilon(D^2 + k^2)\psi + t_{12} \\
\overline{s}_{22} &=& -\overline{p} - 2\epsilon ik^2D\psi 
                                + (1+a)^{-1}k(t_{22}-t_{11}),
\end{eqnarray}
and for the polymer extra stress:
\begin{equation}
(-ik\overline{\omega}+ikU+1)t_{22} = 2aik(T_{11}-1)D\psi + 2ak^2\psi T_{12}
\end{equation}
\begin{multline}                  
(-ik\overline{\omega}+ikU+1)t_{11} = 2ikD\psi + (1-a^2)\gd t_{12} \\ \mbox{}
+ T_{12}((1-a^2)D^2\psi - (1+a^2)k^2\psi)
\end{multline}
\begin{equation}                                   
(-ik\overline{\omega}+ikU+1)t_{12} = (1-T_{11})(D^2+k^2)\psi - \gd t_{11} 
+ 2(1-a^2)^{-1}T_{11}k^2\psi .
\end{equation}
If the interface has been displaced from its original position 
$\overline{\eta}=\kappa$ to a new position 
$\overline{\eta}=\kappa + \zeta\exp{(ikx-i\omega t)}$, then the conditions 
of continuity of velocity and traction across the interface give us 
continuity of the following quantities at $y=\kappa$:
\begin{equation}
-ik\psi
\hspace{8mm} 
\gd\zeta + D\psi
\hspace{8mm} 
-ik\zeta S_{11}+ s_{12}
\hspace{8mm} 
\overline{s}_{22}
\end{equation}
in agreement with the results of (\ref{eq:psi_cont}--\ref{eq:s22_cont}) for 
$l\to 0$.

Finally, the condition that the interface should be a material surface, 
equivalent to (\ref{eq:zeta_def}), gives 
\begin{equation}
(-\overline{\omega} + U(\kappa))\zeta = -\psi(\kappa). 
\end{equation}
This system of equations is essentially fourth-order in $\psi$, so the general 
solution within one shear band will contain four unknowns. The no-slip, 
no-penetration boundary conditions on the walls remove two unknowns in each 
band, leaving a total of four unknowns to be determined through these jump 
conditions at the interface. Since there are four jump conditions here, and 
no driving term not proportional to $\psi$, the existence of a non-zero 
solution to the problem is the eigenvalue condition which allows $\omega$ 
to be fixed.

We expand each perturbation quantity as a regular power series in $k$. A 
critical quantity in each fluid phase is the {\it marginal viscosity}, 
that is, the slope of the stress-shear-rate relation $\sigma(\gd)$ at our 
selected shear rate:
\begin{equation}
\overline{\mu} = \frac{\d S_{12}}{\d\gd} 
= \frac{\d}{\d\gd}(\epsilon\gd + T_{12})
= \epsilon + \frac{(1 - 2T_{11})}{(1+(1-a^2)\gd^2)}.
\end{equation}
Using $X_L$ to denote the value of quantity $X$ in the lower shear-rate phase, 
and $X_H$ to denote its value in the higher shear-rate phase, the solution at 
leading order in $k$ is 
\begin{equation}
s_{12} = \mua D^2\psi_L = \mub D^2\psi_H = Ay + B 
\hspace{10mm} s_{11} = s_{22} = -p = iA 
\end{equation}
where $A$ and $B$ are constants. The physical interpretation of this result 
is that, to this order, the perturbation consists of simply displacing the 
interface: a one-dimensional perturbation. If $\omega=0$ (as it is to this 
order in $k$), then the system is still on the ``adiabatic'' constitutive 
curve of figure~\ref{fig:nonmonotone}, but with a modified stress $T_b$. Thus 
the change in stress is 
\begin{equation}
s_{12} = 
\Delta S_{12} = \frac{\d S_{12}}{\d \gd} \Delta\gd = \overline{\mu}D^2\psi,
\end{equation}
since the change in shear rate at this order is $D^2\psi$. 

Integration of $D^2\psi$ subject to zero perturbation velocity on the channel 
walls determines $\psi$ in each phase, and the interfacial continuity 
conditions then give the values of the two constants:
\begin{eqnarray}
A &=& \frac{6\zeta[\mub\kappa^2 - \mua(\kappa-1)^2]\mua\mub(\gd_H-\gd_L)}
{[\mub\kappa^2-\mua(\kappa-1)^2]^2 - 4\mua\mub\kappa(\kappa-1)}, \\
B &=& \frac{2\zeta[\mua(\kappa^3-3\kappa+2)-\mub\kappa^3]\mua\mub(\gd_H-\gd_L)}
{[\mub\kappa^2-\mua(\kappa-1)^2]^2 - 4\mua\mub\kappa(\kappa-1)}.
\end{eqnarray}
The eigenvalue at this order is real, which corresponds to the perturbation 
travelling in the flow direction, and is given by 
\begin{equation}
\omega \sim \gd_L\kappa k + \frac{2\mua\mub\kappa^2(\kappa-1)^2(\gd_H-\gd_L)k}
{[\mub\kappa^2 - \mua(\kappa-1)^2]^2 - 4\mua\mub\kappa(\kappa-1)} + O(k^2).
\end{equation}
The imaginary part of $\omega$ determines the stability or instability of the 
flow ($\mathrm{Im}(\omega)<0$ for stable flow, $\mathrm{Im}(\omega)>0$ for 
unstable flow). This appears at the next order. This problem can still be 
solved analytically, but the result is too unwieldy to reproduce here. 

\begin{figure}
\begin{center}
\resizebox{80mm}{!}{\rotatebox{-90}{\includegraphics{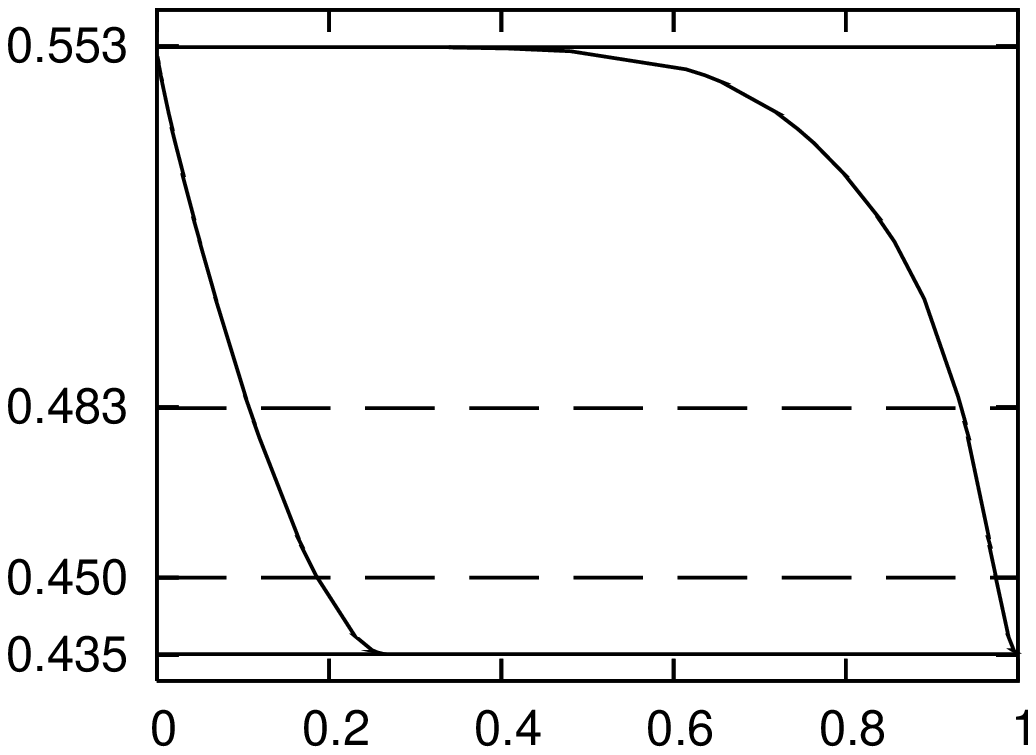}}}
\begin{picture}(0,0)
\put(-110,-65){$U$}\put(-35,-35){$S$}\put(-170,-110){$S$}
\put(-235,-130){\rotatebox{90}{Scaled shear stress,}}
\put(-220,-90){\rotatebox{90}{$(1-a^2)^{1/2}T_b$}}
\put(-190,-160){Proportion of low shear rate band, $\kappa$}
\end{picture}
\end{center}
\caption{\label{fig:longwave}Stability boundaries in parameter space 
for long-wave perturbations (stability or instability at order $k^2$).
The $x$-axis is $\kappa$, the proportion of low-shear-rate fluid in the base 
flow; the $y$-axis is the scaled shear stress $(1-a^2)^{1/2}T_b$. The curves 
shown are for $\epsilon=0.05$, and these boundaries are independent of the 
value of $a$. The solid horizontal lines show the minimum and maximum possible 
values of $(1-a^2)^{1/2}T_b$, being 0.435207 and 0.552806 respectively for 
this value of $\epsilon$. The dotted horizontal lines show the selected values 
of $(1-a^2)^{1/2}T_b$ for the two limiting diffusive models: Olmsted's model 
($\Delta=0$) predicts $(1-a^2)^{1/2}T_b=0.48284$, while Yuan's model 
($\Delta=1$) has $(1-a^2)^{1/2}T_b=0.45$ for $\epsilon=0.05$. $U$ denotes the 
unstable region (typically for two roughly equal shear bands) and $S$ the 
stable regions (typically one or other shear band being very narrow).
The stable regions here have only been shown to be stable to asymptotically 
long waves: as we see in section~\ref{sec:num}, most of these parameters do 
show instability for some value of the wavenumber, $k$.}
\end{figure}

Instead, in figure~\ref{fig:longwave} we show the stability boundaries 
for this long-wave mode of perturbation, plotted in the parameter plane 
of $T_b\sqrt{(1-a^2)}$ against $\kappa$. 

It will be seen that the range of parameter values over which the flow is 
unstable includes most of the available values of $\kappa$, with the exception 
of values giving an interface very close to one of the walls. In particular, 
for the shear stress which is selected by the stress-diffusion model 
$\Delta=0$, there is instability for $0.108<\kappa <0.933$. 

In the next section we give the results of numerical calculations at finite 
values of $k$, and we will show these analytical asymptotes along with those 
numerical results. 

\section{Numerical Results} 
\label{sec:num}

In order to access perturbations with moderate or short wavelengths, we solve 
the linearised equations numerically. The scheme used is a shooting 
method, in which a value of $\omega$ is guessed and the Newton-Raphson method 
is used to find the true value. The equations are integrated inwards from each 
wall and the jump conditions at the interface provide the dispersion relation 
through a determinant condition as introduced by Ho \& Denn~\cite{HoD77}.

\subsection{Most unstable mode}

In this section we choose as illustrative the parameters $\epsilon=0.05$ and 
$a=0.3$, and investigate the wavenumber-dependence of the instability. We use 
the nonlocal model with stress diffusion (\ie\ $\Delta=0$) to select the shear 
stress $T_b=0.506158$ of the one-dimensional base state, but the subsequent 
two-dimensional linear stability calculations are carried out using the local 
Johnson-Segalman model. 

\begin{figure}
\begin{center}
\resizebox{80mm}{!}{\rotatebox{-90}{\includegraphics{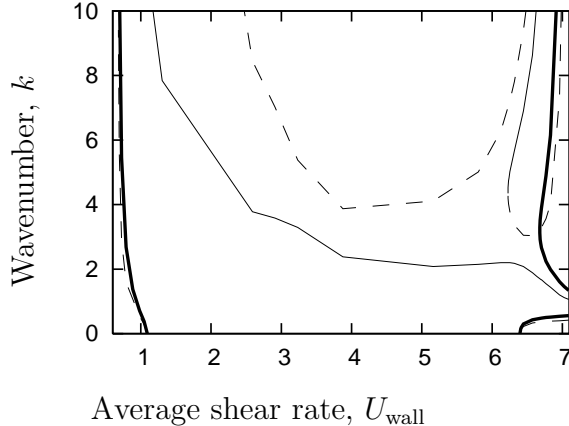}}}
\begin{picture}(0,0)
\put(-230,-115){\rotatebox{90}{Wavenumber, $k$}}
\put(-200,-165){Average shear rate, $U_\mathrm{wall}$}
\end{picture}
\end{center}
\caption{\label{fig:kvskappa}The dependence of the growth rate of instability 
on the average shear rate $U_\mathrm{wall}$ and on wavenumber, $k$.
The parameters are $\epsilon=0.05$, $a=0.3$ and the shear stress is that 
selected by the non-local model with stress diffusion, $T_b=0.506158$. The 
bold solid curves mark the boundaries between instability and stability (the 
unstable region being the centre of the plot), the thin solid lines mark local 
maxima in the plot of growth rate against wavenumber, and the dashed curves 
mark local minima in the same plot.}
\end{figure}

We vary the average shear rate across the channel, $U_\mathrm{wall}$, which 
changes the proportion of the lower shear rate band. Our nondimensionalisation
means that $U_\mathrm{wall}$ is the same as the Weissenberg number. The 
behaviour of the instability against wavenumber of the perturbation changes 
as $U_\mathrm{wall}$ varies. It is only for extremely low shear-rates, that is,
thin high-shear bands that this flow is stable for all wavenumbers. In 
figure~\ref{fig:kvskappa} we plot various curves in the 
$U_\mathrm{wall}$--$k$ plane. The thick solid curves mark the borderline 
between unstable and stable parameters. For most values of $U_\mathrm{wall}$ 
in the shear-banding region $0.661 < U_\mathrm{wall} < 7.089$ these curves 
do not appear as the perturbations at all wavenumbers are unstable. 

The thin solid curves gives the wavenumbers at which the growth rate reaches a 
local peak. Where there is only one such curve (for example, at 
$U_\mathrm{wall}=4$), it is tempting to regard this as the most unstable mode; 
however, within the confines of our local approximation to the diffusive 
Johnson-Segalman fluid, there is an instability to very short waves 
\cite{Ren95}, with a growth rate independent of $U_\mathrm{wall}$, and for 
many layer arrangements this is in fact the most unstable mode. For our 
illustrative parameters, the growth rate of the $k\to\infty$ mode is 0.533, 
and this mode is the most unstable for $U_\mathrm{wall} > 3.2$. 
For lower values of $U_\mathrm{wall}$ the most unstable mode is the peak at 
the lowest wavenumber, although this wavenumber also becomes very large 
as $U_\mathrm{wall}\to \gd_L$. In practice, as we shall 
see in \S\ref{sec:sigma0}, for realistic non-zero values of $l$ we 
expect diffusive effects to damp very short-wave instabilities, and the peak 
growth rate will always occur at finite $k$: in \S\ref{sec:empirical} we will 
show how the maximum growth rate varies with average shear rate, with 
diffusion added in a semi-empirical way. 

\begin{figure}\begin{center}
\scalebox{0.4}{\rotatebox{-90}{\includegraphics{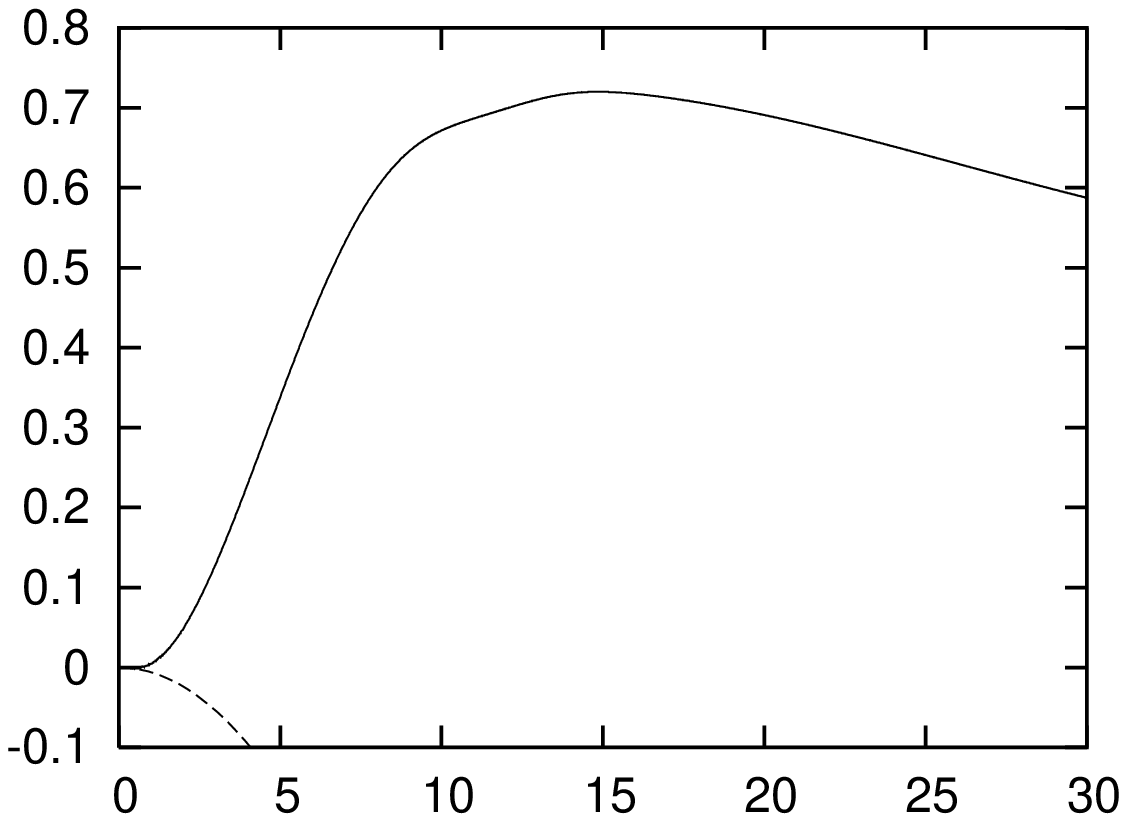}}}\hspace{5mm}
\scalebox{0.4}{\rotatebox{-90}{\includegraphics{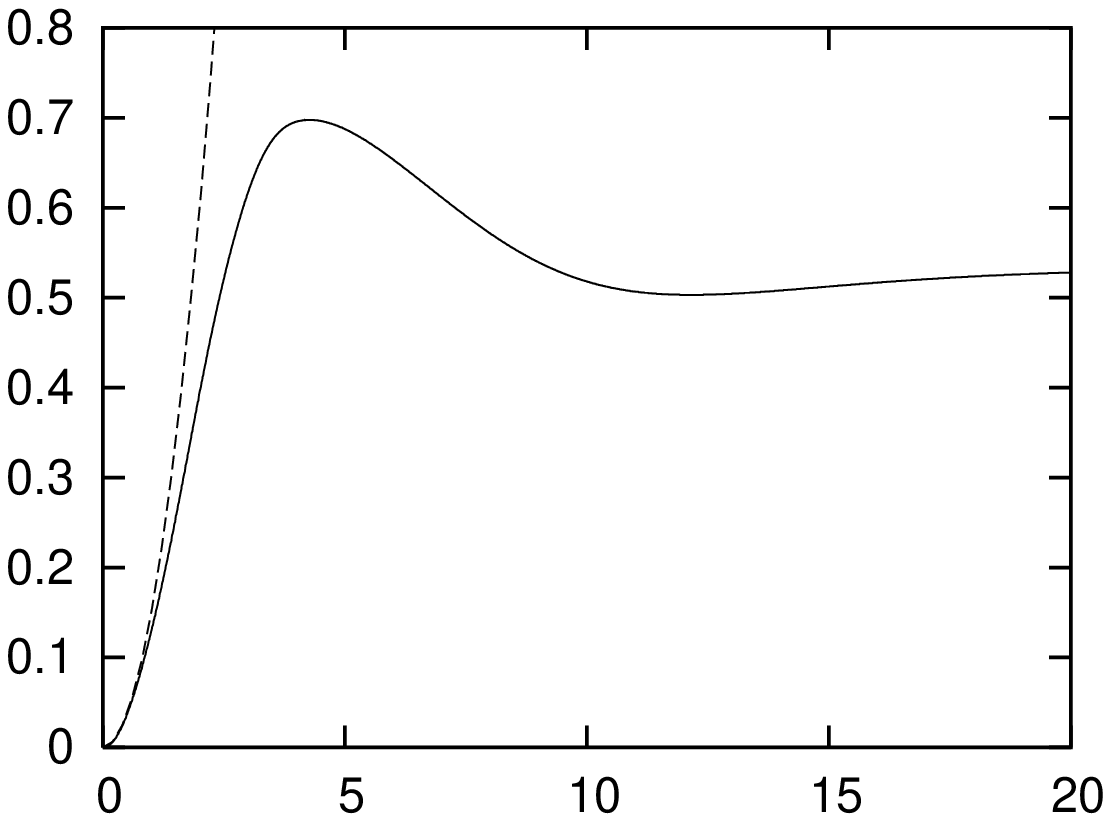}}}\\[8mm]
\scalebox{0.4}{\rotatebox{-90}{\includegraphics{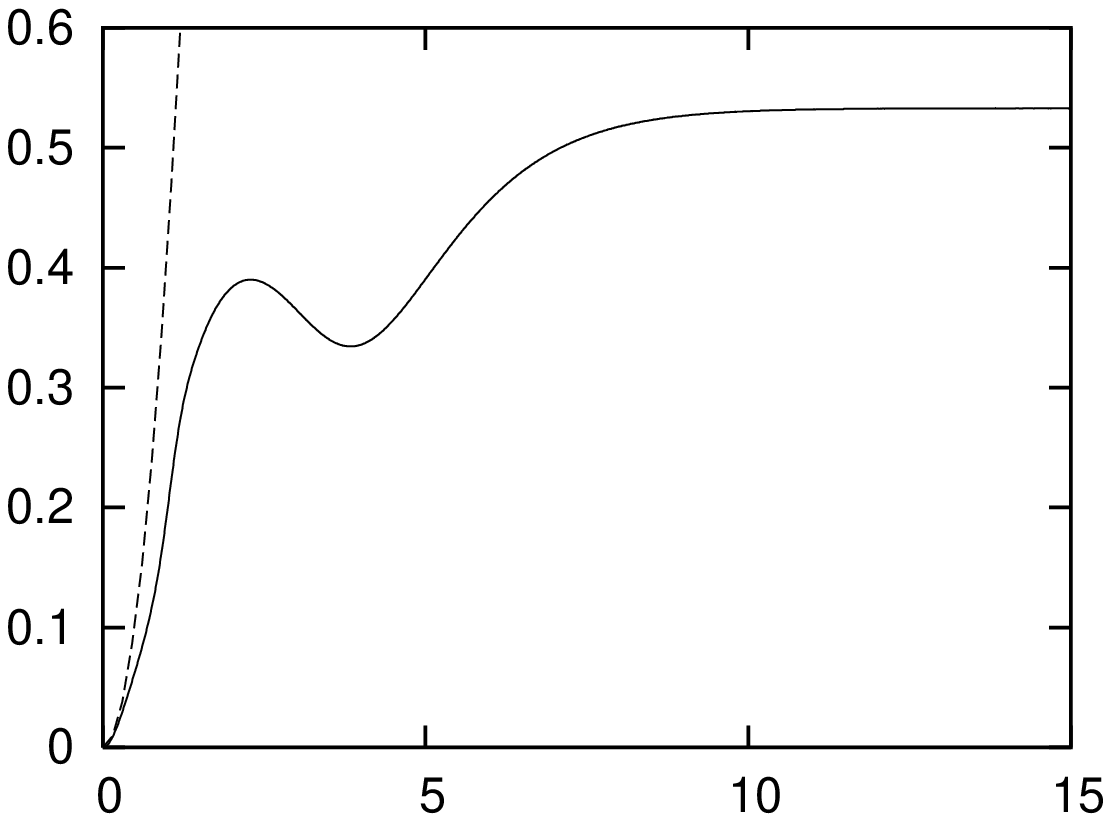}}}\hspace{5mm}
\scalebox{0.4}{\rotatebox{-90}{\includegraphics{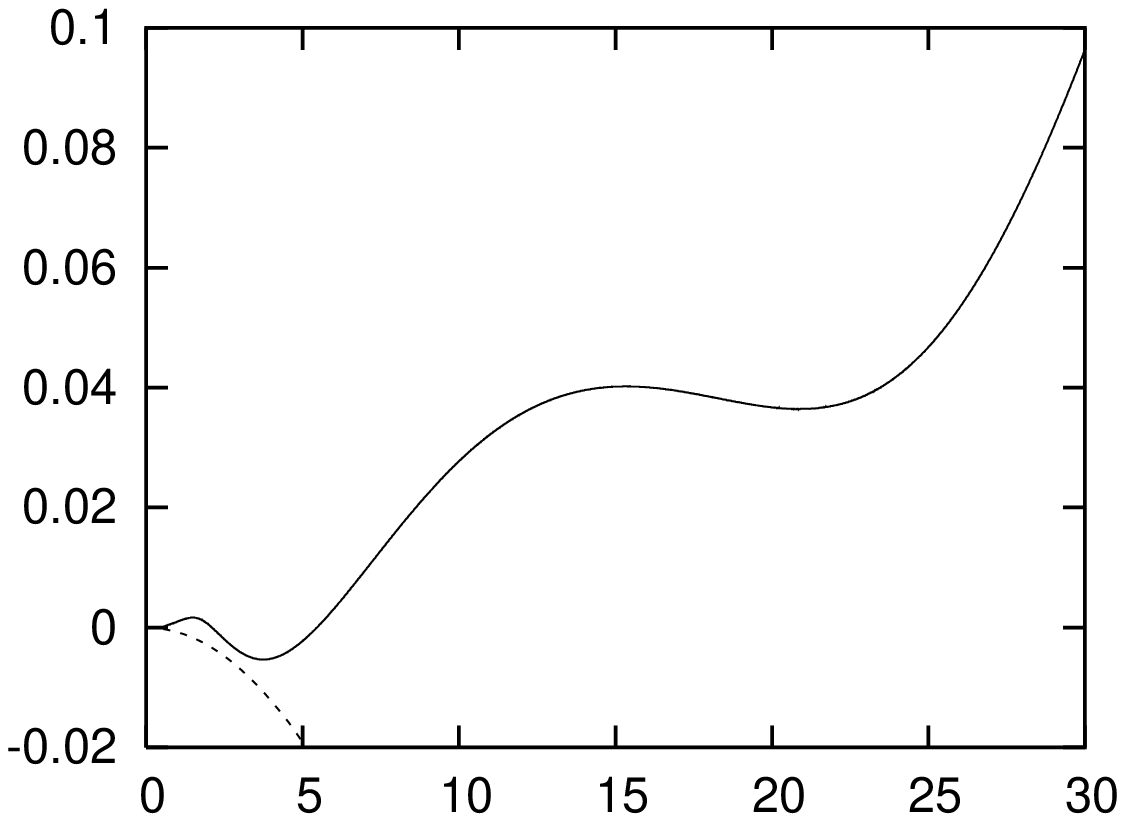}}}\\[5mm]
\begin{picture}(0,0)
\put(-100,167){(a) $U_\mathrm{wall}=1$}
\put(70,165){(b) $U_\mathrm{wall}=2$}
\put(-100,40){(c) $U_\mathrm{wall}=4$}
\put(50,88){(d) $U_\mathrm{wall}=6.8$}
\put(-120,127){Wavenumber, $k$}\put(50,127){Wavenumber, $k$}
\put(-120,0){Wavenumber, $k$}\put(50,0){Wavenumber, $k$}
\put(-160,140){\rotatebox{90}{Growth rate, $\mathrm{Im}(\omega)$}}
\put(-160,18){\rotatebox{90}{Growth rate, $\mathrm{Im}(\omega)$}}
\put(0,140){\rotatebox{90}{Growth rate, $\mathrm{Im}(\omega)$}}
\put(0,18){\rotatebox{90}{Growth rate, $\mathrm{Im}(\omega)$}}
\end{picture}
\end{center}
\caption{\label{fig:l=0plots}Plots of growth rate against wavenumber for four 
different values of $U_\mathrm{wall}$, the average shear rate. The parameters 
here are $\epsilon=0.05$, $a=0.3$, $l=0$, and $T_b=0.506158$, the value 
predicted by the stress-diffusion model. In each case the order $k^2$ 
behaviour of long waves is a dotted curve.
(a) $U_\mathrm{wall}=1$. 
Long waves are stable with decay rate $\mathrm{Im}(\omega) \sim -0.0060k^2$ 
and instability begins at $k>0.65$ (not discernible on the scale of the plot). 
The region $k>30$, in which the growth rate tends smoothly to 0.533 from above,
is omitted to allow a clearer view of the behaviour for longer waves.
(b) $U_\mathrm{wall}=2$. 
Long waves are unstable with growth rate $\mathrm{Im}(\omega) \sim 0.15k^2$, 
and the most unstable mode is at $k=4.3$ with growth rate $0.698$.
(c) $U_\mathrm{wall}=4$. 
Long waves are unstable with growth rate $\mathrm{Im}(\omega) \sim 0.42k^2$, 
and there is a peak in growth rate at $k=2.3$, but the most unstable mode is 
$k\to\infty$. 
(d) $U_\mathrm{wall}=6.8$.
Here, as in (a), asymptotically long waves are stable. Long waves have 
decay rate $\mathrm{Im}(\omega) \sim -0.00076k^2$. We have omitted the region 
$k>30$, in which the growth rate tends smoothly to 0.533 from below.}
\end{figure}

In figure~\ref{fig:l=0plots} we give four plots of growth rate against 
wavenumber, to illustrate the different types of behaviour seen in 
figure~\ref{fig:kvskappa}. The parameters chosen are $U_\mathrm{wall}=1$, for 
which long waves are stable, and instability arises at finite $k$; 
$U_\mathrm{wall}=2$, which has one maximum which is the most unstable mode; 
$U_\mathrm{wall}=4$, which has a single local maximum but for which the most 
unstable mode is for asymptotically short waves; and $U_\mathrm{wall}=6.8$, 
for which (again) long waves are stable, and two separate unstable regions 
$0.51<k<2.1$ and $k>5.45$ are observed. In each case the behaviour as 
$k\to\infty$ is the short-wave instability predicted by Renardy~\cite{Ren95}, 
with growth rate 0.533, and the long-wave behaviour matches the analytic 
calculation of section~\ref{sec:longwave}. The long-wave asymptotes are 
plotted along with the numerical calculations.

\subsection{Comparison with results at small finite $l$; Fielding (2005)}
\label{sec:sigma0}

In this section we carry out a comparison with the numerical study of the full
diffusive Johnson-Segalman model at $\Delta=0$, first published in 
\cite{Fie05}. From the findings of section~\ref{sec:lowl}, we expect that, 
for a given one-dimensional base state, the numerical results of \cite{Fie05}
should converge, in the limit $l\to 0$, to the asymptotic results calculated 
here at $l=0$. 

\begin{figure}\begin{center}
\resizebox{60mm}{!}{\rotatebox{-90}{\includegraphics{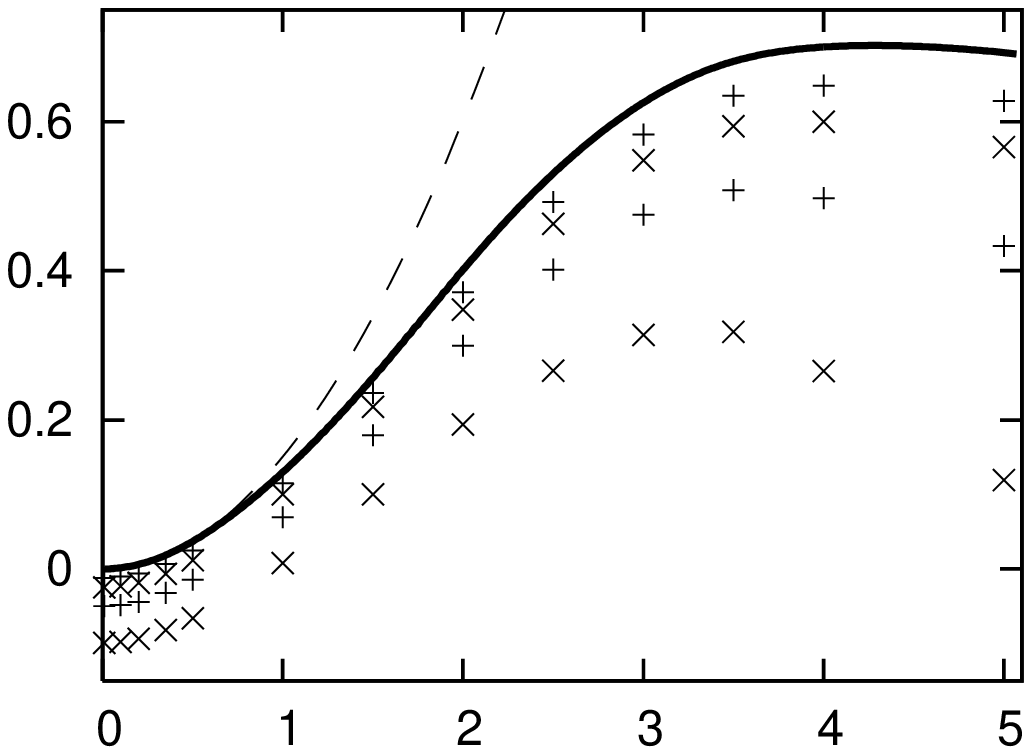}}}
\hspace{4mm}
\begin{picture}(0,0)
\multiput(-140,-120)(210,0){2}{Wavenumber, $k$}
\multiput(-185,-110)(210,0){2}{\rotatebox{90}{Growth rate, $\mathrm{Im}(\omega)$}}
\end{picture}
\hspace{4mm}
\resizebox{60mm}{!}{\rotatebox{-90}{\includegraphics{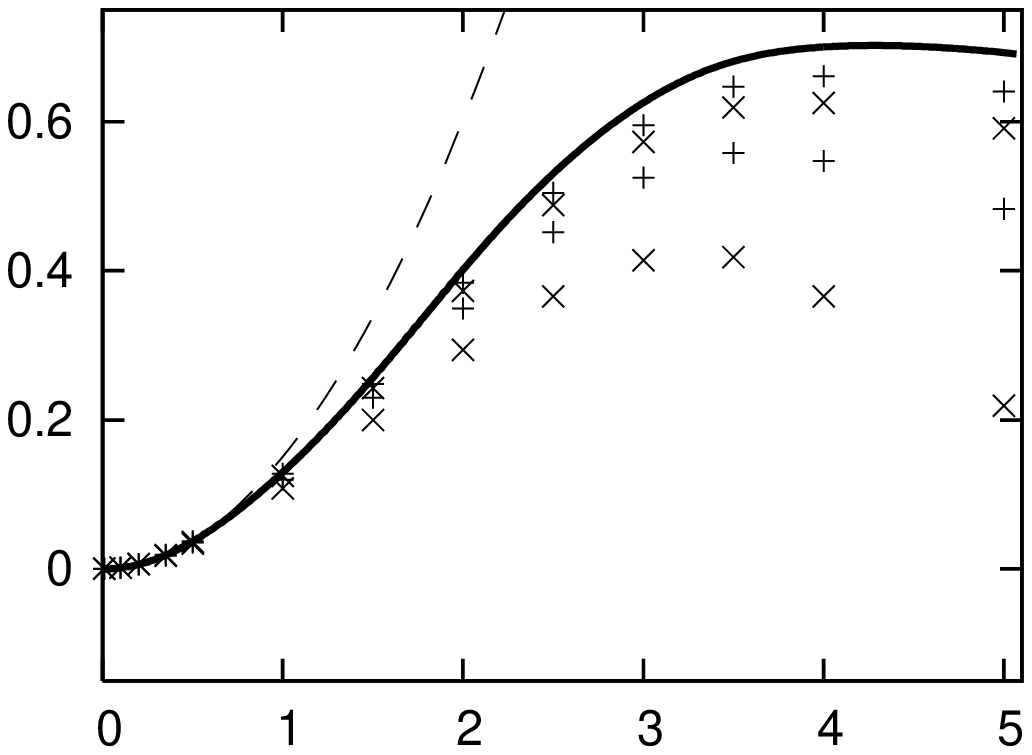}}}
\end{center}
\caption{\label{fig:suzannefig3}Growth rate $\mathrm{Im}(\omega)$ plotted 
against wavenumber, $k$ for Couette flow at $\epsilon=0.05$, $a=0.3$, 
$U_\mathrm{wall}=2$, $T_b=0.506158$. Curves: $l=0$, calculation for a 
non-diffusive Johnson-Segalman fluid with no material transport across the 
interface between phases. The dashed curve is the long-wave asymptotic form 
$\mathrm{Im}(\omega)\sim 0.1506k^2$; the solid curve is the full numerical 
calculation. Points: (from highest to lowest) $l=0.00125$, 0.0025, 0.005, 0.01.
The figure on the left shows the true growth rate; on the right for the 
finite-$l$ results we have plotted $\mathrm{Im}(\omega) + 10l$.}
\end{figure}

In figure~\ref{fig:suzannefig3}, therefore, we reproduce (as points) the data 
from Figure 3 of \cite{Fie05} showing instability for small finite values of 
$l$, along with (as curves) numerical results for the pure, $l=0$ 
Johnson-Segalman fluid under the assumption that the interface is a material 
surface, and the long-wave asymptotic form for the same $l=0$ situation. 
On the left we present the raw data. From 
the numerical data we observe that $\mathrm{Im}(\omega) \sim ak^2 + b(l)$ 
for long waves $k < 1$, with $b(l)\approx-10l$. We confirm the scaling of 
$b(l)$ with $l$ in the appendix. On the right, therefore, we add an additional 
term of $10l$ to the numerical results for finite $l$, and observe that for 
long waves this collapses all the points onto the $l=0$ curve. 

\begin{figure}
\begin{center}
\resizebox{80mm}{!}{\rotatebox{-90}{\includegraphics{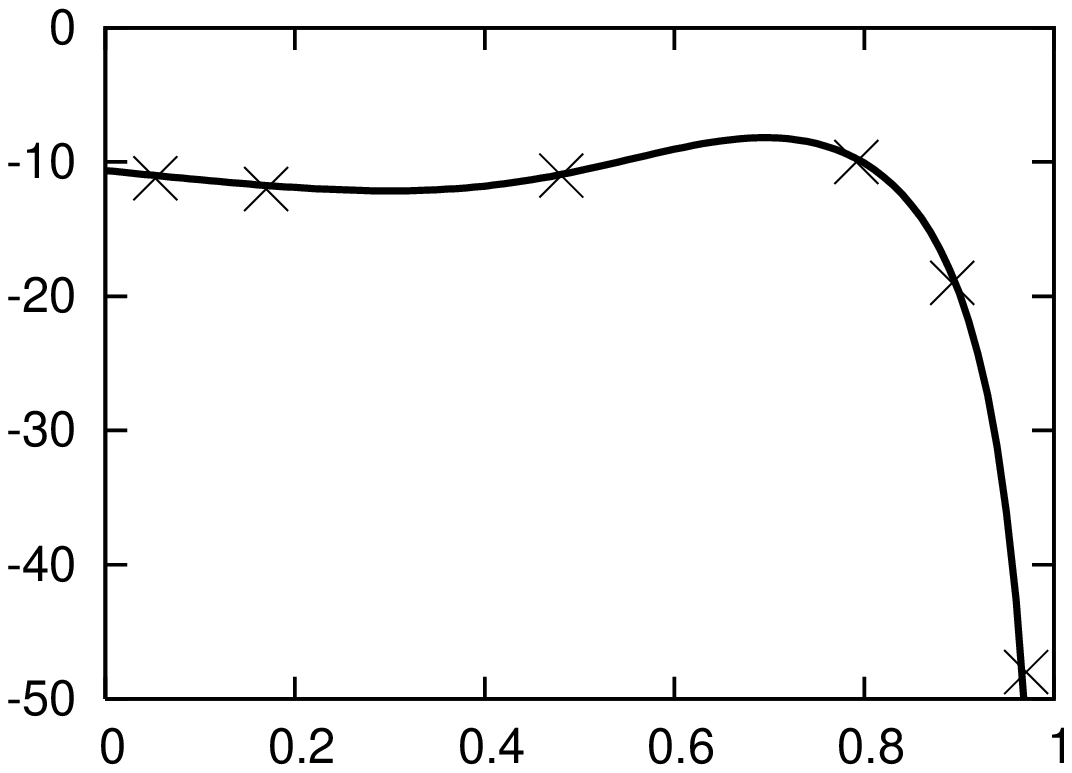}}}
\begin{picture}(0,0)
\put(-230,-145){\rotatebox{90}{Coefficient of $l$ at $k=0$: $\sigma_0$}}
\put(-190,-160){Proportion of low shear rate band, $\kappa$}
\end{picture}
\end{center}
\caption{\label{fig:sigma0} Dependence of the coefficient of $l$ in the 
growth rate of very long-wave perturbations on the interface position, 
$\kappa$. Here $a=0.3$, $\epsilon=0.05$ and $T_b=0.506158$. The points are 
from numerical calculations and the curve is given by the analytical 
prediction from (\ref{eq:sigma0}), with $\tilde{\sigma}$ chosen to match the 
data at $\kappa=0.79$.}
\end{figure}

The numerical factor $10$ which is used at order $l$ for the long-wave results 
is not calculated analytically, but deduced from the $k\to 0$ intercepts from 
the data for the different values of $l$. However, we can make some progress 
towards calculating this value. A full derivation of this process is given in 
appendix~\ref{app:lowklowl}: the final conclusion is that, as $k\to 0$ and 
$l\to 0$, the growth rate can be written as 
$\omega = i\sigma_0 l + O(l^2) + O(k)$, where 
\begin{equation}
\sigma_0 \sim 
\frac{-4[\mub\kappa^3- \mua(\kappa-1)^3]\mua\mub(\gd_H-\gd_L)\tilde{\sigma}}
{[\mub\kappa^2 - \mua(\kappa-1)^2]^2 - 4\mua\mub\kappa(\kappa-1)} 
\label{eq:sigma0}
\end{equation}
and $\tilde{\sigma}$ is an unknown parameter which depends on $a$, $\epsilon$ 
and $T_b$ but not on $U_\mathrm{wall}$ or $\kappa$. For the values of $a$, 
$\epsilon$ and $T_b$ associated with figure~\ref{fig:suzannefig3} the data 
suggest that $\tilde{\sigma}\approx 14$. In figure~\ref{fig:sigma0} we plot 
a numerical calculation of $\sigma_0$ against $\kappa$ for these values of 
$a$, $\epsilon$ and $T_b$. The curve is given by the prediction of 
(\ref{eq:sigma0}) with $\tilde{\sigma}=14$. 
The agreement between theory and numerical calculation of this term is 
remarkable. As $\kappa\to 1$ our prediction (with $\tilde{\sigma}=14$) is 
$\sigma_0 \approx -130$: numerically it is not possible to investigate 
extremely narrow bands as the diffusive layer needs to be clear of the walls, 
but the numerical calculations of this long wave intercept match the 
theoretical prediction well even at $\kappa=0.97$ where $\sigma_0 \approx -48$.

This diffusion-induced stability to very long waves $k\to 0$ was to be 
expected. Calculations had already shown that an interface between shear 
bands at this selected shear stress should be stable to 
one-dimensional perturbations: that is, if the whole interface is rigidly 
displaced from the selected position it should relax back there. A simple 
displacement of the interface corresponds to the limit of very long waves, 
and so this exponential relaxation is precisely the negative growth rate we 
have calculated in the limit $k\to 0$ with $l$ small but finite. 

\begin{figure}
\begin{center}
\resizebox{80mm}{!}{\rotatebox{-90}{\includegraphics{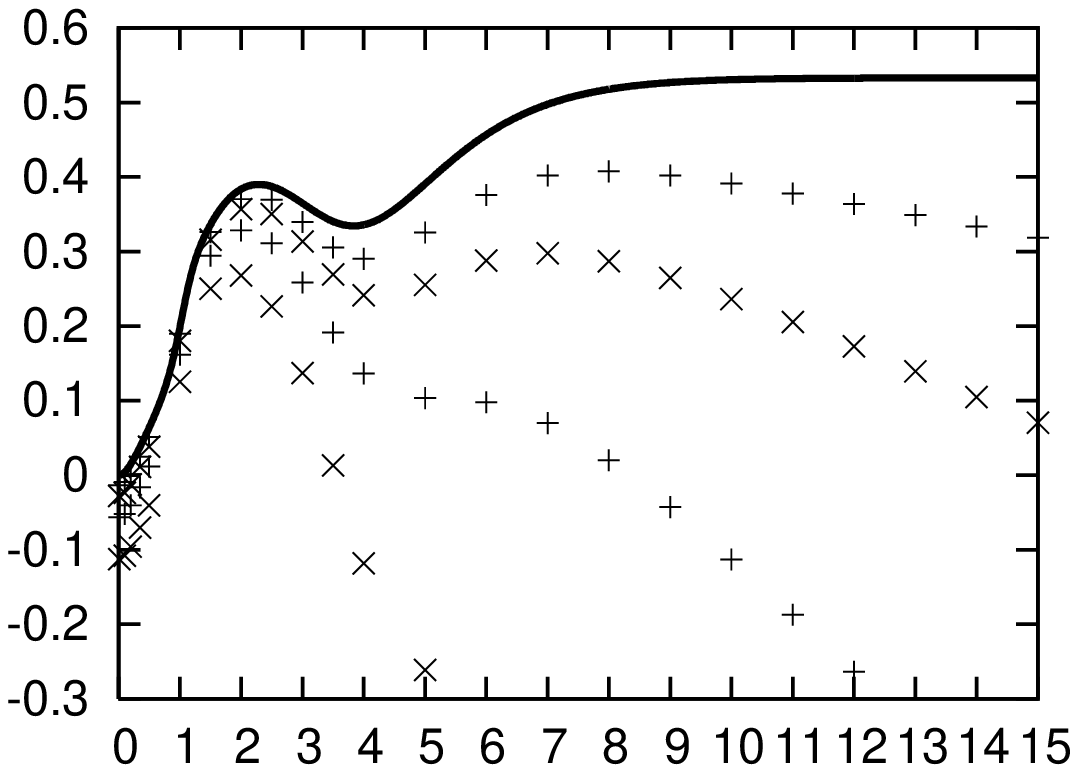}}}
\begin{picture}(0,0)
\put(-230,-115){\rotatebox{90}{Growth rate, $\mathrm{Im}(\omega)$}}
\put(-150,-160){Wavenumber, $k$}
\end{picture}
\end{center}
\caption{\label{fig:kappa0.4}Dependence of the growth rate of the instability 
on wavenumber. Parameters $\epsilon=0.05$, $a=0.3$, $T_b=0.506158$, 
$U_\mathrm{wall}=4$. The long-wave mode leads to 
a peak growth rate at $k\approx 2$, but the growth rate increases again for 
shorter waves, and the most unstable mode (in the limit $l=0$) is for very 
short waves. The solid curve is for $l=0$; the points (from highest to lowest)
are $l=0.00125$, 0.0025, 0.005, 0.01.}
\end{figure}

In figure~\ref{fig:kappa0.4} we show another comparison between the $l=0$ and 
$l\not=0$, $\Delta=0$ cases, this time at $U_\mathrm{wall}=4$. The convergence 
of the $l\not=0$ results to the $l=0$ case as $l\to 0$ is clearly visible. 
At $l=0$ the growth rate for long waves is $\mathrm{Im}(\omega) \sim 0.42k^2$, 
and the calculation of section~\ref{sec:sigma0} predicts $\sigma_0 = -10.9$ 
for these parameters, so for long waves at small finite $l$ we expect 
$\mathrm{Im}(\omega) \sim -10.9l + 0.42k^2$, which means that instability 
first appears at $k \approx 5.1l^{1/2}$. This scaling $k \sim l^{1/2}$ for 
the lowest unstable wavenumber is universal provided that long waves are 
unstable in the $l=0$ case. 

For very short waves, we can see from figure~\ref{fig:kappa0.4} that the 
addition of diffusion has a large effect on the eigenvalue. We expected this 
when we stated in section~\ref{sec:analysis-small-l} that our analysis would 
only be valid for $k \ll l^{-1}$. In fact we can see empirically that for 
these parameters, $\mathrm{Im}(\omega) \sim -12kl + O(1)$ for fixed $l$ as 
$k\to\infty$, leading us to predict instability for $k \ll l^{-1}$ and 
stability for very large $k$. The size of the prefactor in this case means 
that the results for finite $l$ deviate from the $l=0$ limit earlier (as $k$ 
increases) than a simple scaling argument might have led us to expect. 

In summary, if $0.108 < \kappa < 0.933$ (for the stress-diffusion model) then 
in the limit $l=0$ the flow is unstable to perturbations of all wavenumbers, 
and in this case we expect the diffusive flow to be unstable over a large 
region $l^{1/2} \ll k \ll l^{-1}$. 

\begin{figure}
\begin{center}
\resizebox{80mm}{!}{\rotatebox{-90}{\includegraphics{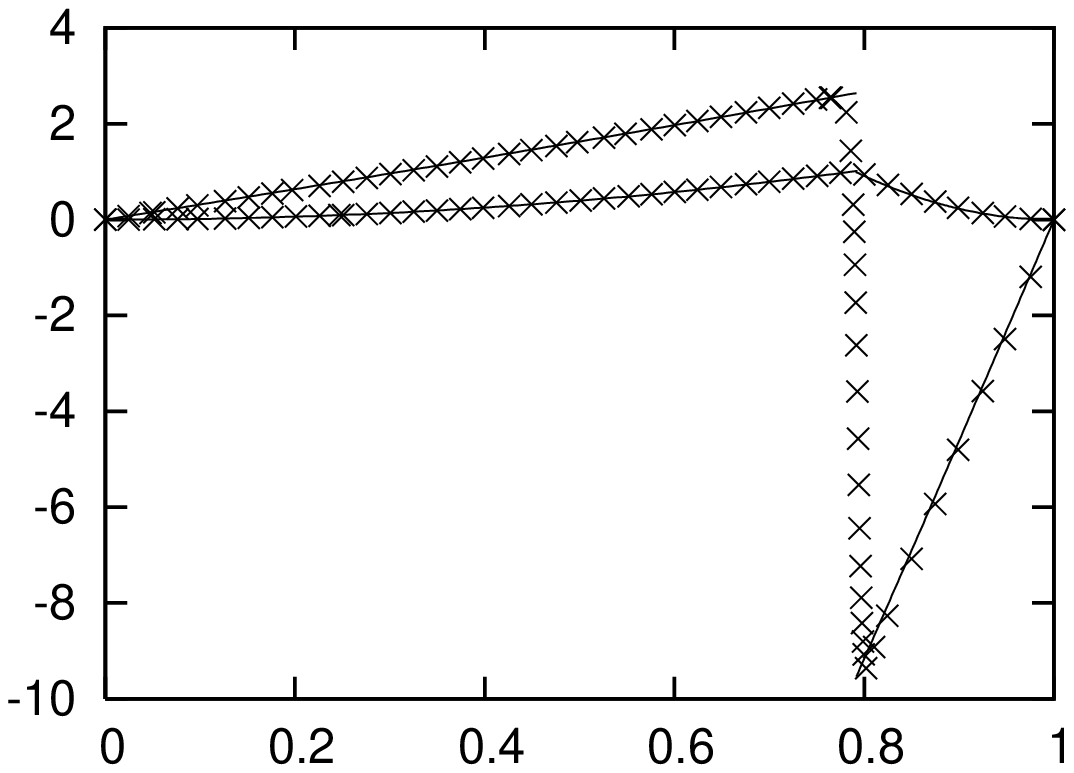}}}
\begin{picture}(0,0)
\put(-230,-140){\rotatebox{90}{Streamfunction $\psi$ and $D\psi$}}
\put(-170,-160){Channel position, $y$}
\end{picture}
\end{center}
\caption{\label{fig:stream}The streamfunction $\psi$, proportional to $v$ 
(continuous function) and its derivative $D\psi$, or $u$ (discontinuous) of 
the least stable mode at $\epsilon=0.05$, $a=0.3$, $U_\mathrm{wall}=2$, 
$T_b=0.506158$, $k=0.1$. This mode is unstable at $l=0$, with growth rate 
0.00148; at $l=0.005$ it is stable with growth rate -0.0485454. The crosses 
are from the full calculation at $l=0.005$, and the solid lines (lying under 
the crosses except near the interface) from the limit $l=0$ using the material 
surface condition. The streamfunction is plotted against position across the 
channel, $y$, and is normalised such that $\psi(\kappa)=1$. We show only the 
real part of $\psi$ and $D\psi$ here; the imaginary part is smaller by a 
factor of order $k$.}
\end{figure}

As a final comparison between our calculations and the numerics for small 
finite $l$, in figure~\ref{fig:stream} we give the perturbation streamfunction 
$\psi$ and its first derivative (proportional respectively to the cross- and 
along-channel velocity components) for 
one specific mode. We have plotted the real part of $\psi$ and its derivative 
at the phase ($x$-position) where $\psi(\kappa)$ is real. The parameters 
(given in the caption to figure~\ref{fig:stream}) are such that, in the limit 
$l=0$, long waves are unstable and the mode $k=0.1$ shows this growth. 
However, for $l=0.005$ the decay term at $O(l)$ dominates and the mode is 
stable. Nonetheless, the form of the streamfunction (equivalent to $10i$ times 
the $y$-velocity in this case) and its first derivative (which is the 
$x$-velocity) is extremely similar between the two modes.

\subsubsection{Most unstable mode with diffusion} 
\label{sec:empirical}

\begin{figure}
\begin{center}
\resizebox{80mm}{!}{\rotatebox{-90}{\includegraphics{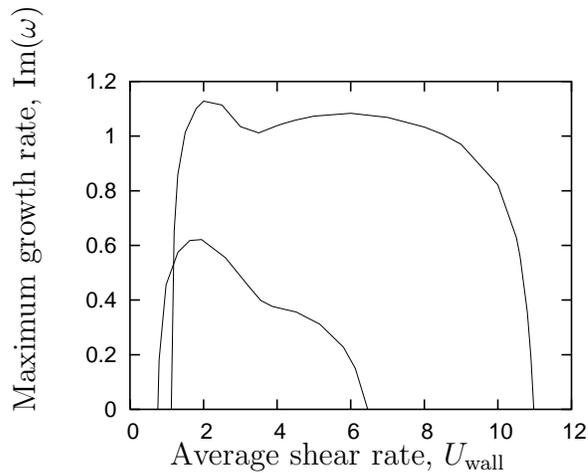}}}
\begin{picture}(0,0)
\put(-230,-135){\rotatebox{90}{Maximum growth rate, $\mathrm{Im}(\omega)$}}
\put(-170,-155){Average shear rate, $U_\mathrm{wall}$}
\end{picture}
\end{center}
\caption{\label{fig:empirical}Maximum growth rate plotted against average 
shear rate for a semi-empirical model of the stress-diffusion fluid with 
$l=0.00125$, $\epsilon=0.05$, and $\Delta=0$. Lower curve $a=0.3$; 
upper curve $a=0.8$. In both cases r\'egimes with very narrow bands of one 
or other shear rate are stable to all perturbations. The wavenumber of the 
most unstable waves increases as we approach these stable regions.}
\end{figure}
Finally, in figure~\ref{fig:empirical} we give an empirical idea of the 
maximum growth rates which might be expected to be seen for a range of 
average shear rates, and for two different values of the slip parameter $a$. 
For the pure JS model, the most unstable mode is often the short-wave limit 
$k\to\infty$, which will be stabilised by diffusion for the modified model. 
It is therefore unhelpful to give the growth rate of the most unstable mode; 
rather, we attempt to give the expected most unstable growth rate in the 
presence of some diffusion $l=0.00125$. In order to carry out the computations 
for a large variety of parameters, we make two very broad assumptions: 
\begin{itemize}
\item Based on curve fitting of the data in figure~\ref{fig:kappa0.4}, the 
growth rate at any $k$ and any $a$ may be reasonably approximated by 
\[ \mathrm{Im}(\omega)_\mathrm{approx} = \mathrm{Im}(\omega)_\mathrm{JS} 
+ \sigma_0 l - 13kl. \]
\item In determining $\sigma_0$, the value $\tilde{\sigma}$ in 
equation~(\ref{eq:sigma0}), which is a function of $a$, $\epsilon$ and $T_b$, 
will be taken to be $14$ (the true value at $a=0.3$, $\epsilon=0.05$ and 
$T_b=0.505158$) independent of these parameters.
\end{itemize}
Using these two assumptions and the numerical calculations for the pure JS 
problem, we predict the wavenumber and growth rate of the most unstable mode 
for two different fluids. We use the stress-diffusion model to select the 
shear stress in each case, and consider the cases $\epsilon=0.05$, $a=0.3$ 
and $\epsilon=0.05$, $a=0.8$. 

There are configurations which are stable to all perturbations, which are 
those layer arrangements with one or other shear band being very narrow, 
\ie\ $U_\mathrm{wall} \approx \gd_L$ or $U_\mathrm{wall} \approx \gd_H$. 
However, the vast majority of mean shear-rates in the shear banding r\'egime 
produce banded flows which are linearly unstable with a moderate growth rate.
The figure suggests that influence of the larger slip parameter $a=0.8$ tends 
to enhance the instability, but we cannot draw concrete conclusions from such 
an empirical model.

\subsection{Comparison with other previous studies} 

\subsubsection{McLeish (1987)}

An early paper by McLeish~\cite{McL87} considered capillary flow with a 
constitutive equation with a non-monotonic flow curve. He predicted exactly 
the opposite of the long-wave behaviour we have found: for slow (\ie\ 
long-wave) perturbations, he found instability only for very narrow regions of 
high-shear-rate material (the band close to the wall). That work used a 
slightly different constitutive equation~\cite{McL86}, based on reptation 
theory for linear polymers, but his model has similar behaviour to ours in 
steady shear. He described the stability property of the flow as a dependence 
of the throughput in the lower-shear rate region on both the absolute position 
and the slope of the interface, with a formulation which is mathematically 
equivalent to ours except for the different base flow.

In terms of instability mechanism, McLeish suggests that since a normal stress 
difference is required in order for the system to ``see'' the gradient of the 
interface perturbation, normal stress effects are critical to the instability.
From our calculations we see that this does appear to be true, but that the 
mechanism of instability is not quite the clean recirculation mechanism 
found by Hinch \ea~\cite{Hin92} for coextruded fluids having matched 
viscosities and a jump in $N_1$ across the interface between them. In our 
equations there are two driving terms: the jump in $D\psi$, proportional to 
the difference in the base-state shear rate across the interface and the 
interface displacement (\ref{eq:Dpsi_jump}); and the jump in $s_{12}$, 
proportional to the difference in the base-state $N_1$ across the interface 
and the slope of the interface (\ref{eq:s12_jump}). Algebraically, we can 
artificially separate these out, and in most cases studied here the normal 
stress term was weakly stabilising, and the instability comes from the 
interaction of the shear-rate-jump term with the normal stresses in the bulk 
of each fluid.

\subsubsection{Renardy (1995)}
\label{sec:renardy}

Renardy~\cite{Ren95} examined the stability of the local JS model in
planar banded Couette flow.  She found linear instability for short
wavelengths (wavenumber greater than 8). For mainly historical
reasons, however, she happened to confine her study to a base state
corresponding to ``top-jumping'' ($T_b=T_2$). 

\begin{figure}
\begin{center}
\resizebox{80mm}{!}{\rotatebox{-90}{\includegraphics{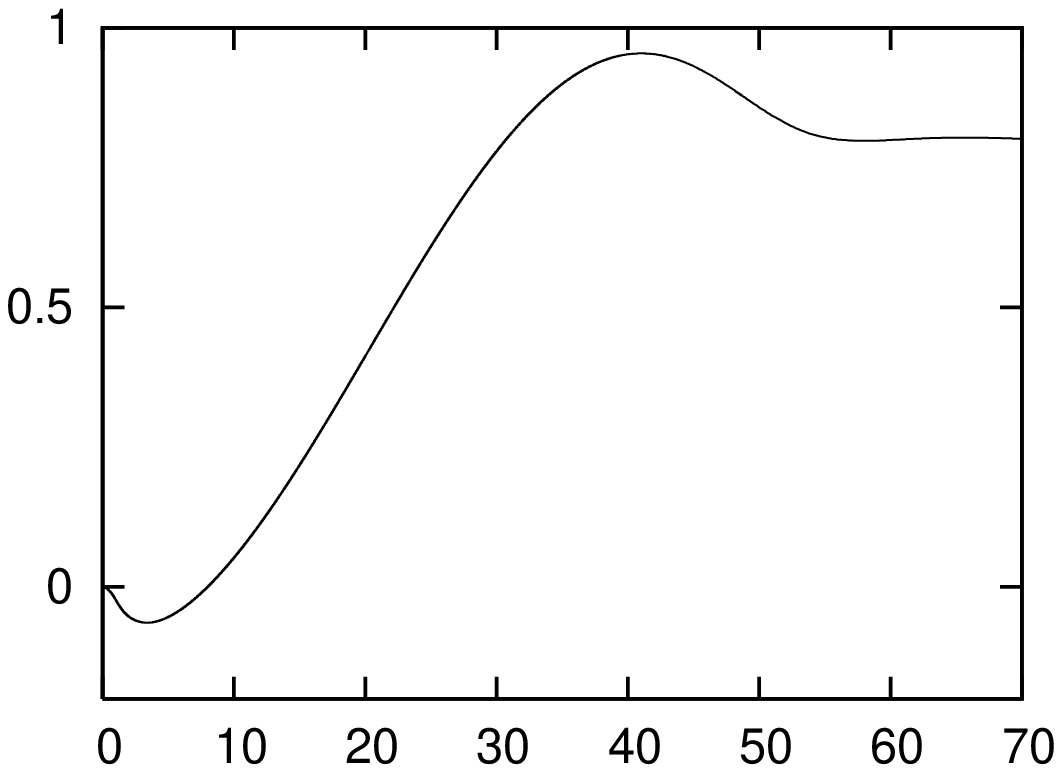}}}
\\[5mm]
\resizebox{80mm}{!}{\rotatebox{-90}{\includegraphics{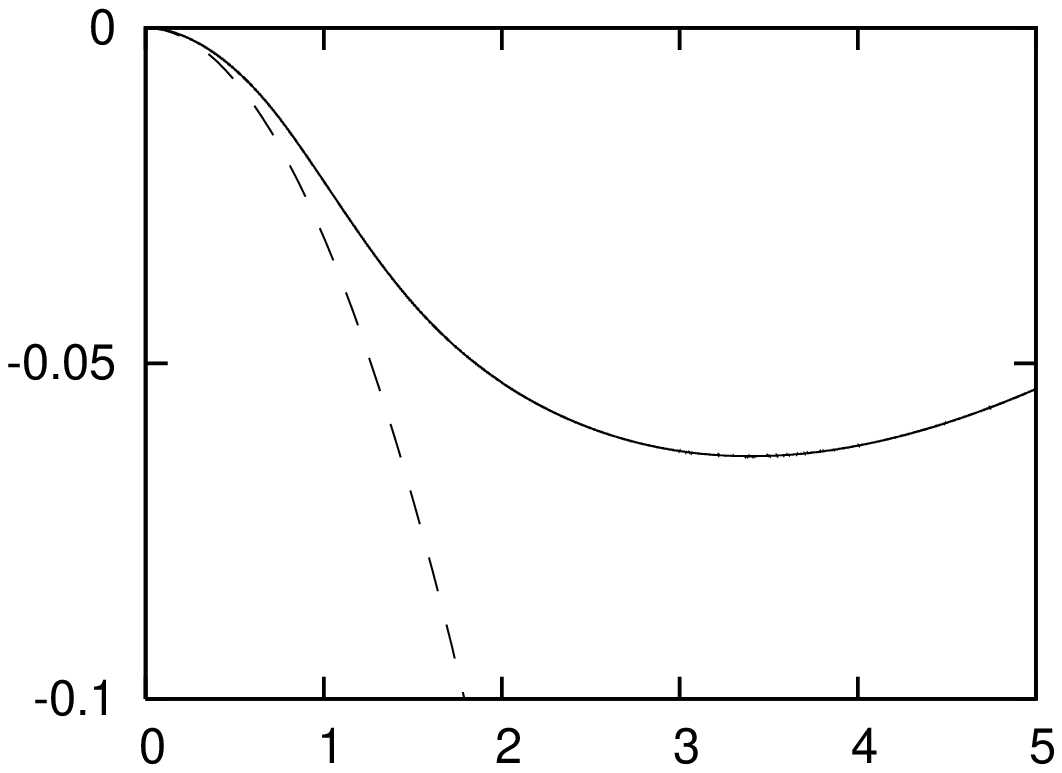}}}
\begin{picture}(0,0)
\put(-230,-125){\rotatebox{90}{Growth rate, $\mathrm{Im}(\omega)$}}
\put(-140,-160){Wavenumber, $k$}
\put(-230,50){\rotatebox{90}{Growth rate, $\mathrm{Im}(\omega)$}}
\put(-140,15){Wavenumber, $k$}
\end{picture}
\end{center}
\caption{\label{fig:renardyfig2}Reproduction of figure 2 from 
Renardy~\cite{Ren95}. Parameters are $\epsilon=0.05$, $a=0.8$, $T_b=0.921$. 
Growth rate is plotted against wavenumber. The upper figure reproduces the 
original exactly (to the naked eye; the original data were not available) and
in the lower figure we reproduce the long-wave portion of the graph. 
Solid curve: numerical calculations; dotted line: asymptote $-0.0312k^2$ from 
long wave calculation. As stated in the text of \cite{Ren95}, instability 
appears for waves shorter than $k\approx 8$, and the peak growth rate of 
instability is for waves having $k\approx41$.}
\end{figure}

As a check on both our analysis and our numerical eigenvalue calculation, we 
reproduce figure 2 that paper. In Figure~\ref{fig:renardyfig2} 
we show our own numerical calculation, which duplicates her results as far as 
can be seen from the graph in \cite{Ren95}. In the second part of the 
figure we plot the same growth rate again, along with the long-wave asymptotic 
form for the growth rate as it depends on wavenumber. We have restricted the 
scale in this second graph in order to better see the accuracy of the 
long-wave result.

Since neither of the two non-local models predicts the top-jumping stress 
$T_2$ as the selected stress, this work is unlikely to be directly relevant 
physically; moreover, the bulk of \cite{Ren95} focuses on a short-wave 
instability. While this instability does occur for the original JS model 
systems at the true stress selected by a nonlocal model 
(for instance, at $\epsilon=0.05$ and $a=0.3$ using $T_b$ as selected by 
stress diffusion, it has growth rate 0.533), when diffusion is added the 
short-wave instability mechanism is destroyed and this mode will not be seen.

\subsubsection{Yuan (1999)}
\label{sec:yuan}

In 1999, Yuan~\cite{Yua99} carried out time-dependent simulations using a 
model corresponding to $\Delta=1$ in our model. He found that, for 
$\epsilon=0.05$ and $a=0.8$, the system uniquely selected a shear stress of 
$T_b=0.81\pm0.04$ for any average shear rate in the range over which a 
homogeneous solution would be on the unphysical descending branch of the 
constitutive curve. This gives values $T_b^*=(1-a^2)^{1/2}T_b=0.486\pm0.024$, 
which is slightly outside the true value of $0.45$ from 
equation~(\ref{eq:yuanTb}). 

He found stable steady states for all parameters in this range. Our long-wave 
analysis predicts instability to long waves at his parameter values for 
$0.186 < \kappa < 0.974$, that is, $1.13 < U_\mathrm{wall} < 7.56$, suggesting 
that several of his simulated flows (at $U_\mathrm{wall}=2$, 3, 4, 5 and 6)
should be unstable to long-wave perturbations. However, he was simulating 
a finite length of channel (and with a relatively large value of $l$) which 
allows access only to specific wavenumbers, which may be the reason that he 
did not capture the instability we have demonstrated. Alternatively, Yuan's 
results may be an artefact of averaging over time $t$ and/or flow location, 
$x$. Note that if these simulations are time-averages of a nonlinearly 
fluctuating interface where a flat interface is linearly unstable, then we 
would expect the average shear stress reported to be higher than the selected 
value for the 1D base state, as observed above.

Another comment which seems unusual is that Yuan states that even when $l=0$, 
the interface has finite width. Although there are physical reasons why this 
may be true in practice, it is not predicted by the governing equations in the 
limit $l=0$: perhaps his observation is a grid-scale effect.

\section{Conclusions and Discussion}
\label{sec:conc}

We have investigated the two-dimensional linear stability of plane Couette 
flow of a shear-banding fluid. If the equations governing the Johnson-Segalman 
model are regularised using a small amount of stress diffusion to provide 
a uniquely selected one-dimensional banded base state, we have 
shown that as far as two-dimensional linear stability is concerned, the limit 
of no diffusion is a regular limit in which the interface region becomes a 
strict material interface. 

Using this limit, we have demonstrated that there is a long wave instability 
for almost all possible positions of a shear band. Only the case of a very 
narrow ``spurted'' region of high shear rate is stable to long-wave 
perturbations. This is in contrast to earlier work by McLeish~\cite{McL87}, 
who found (for Poiseuille flow and slightly different constitutive assumptions)
stability except for the case of a narrow high-shear rate band; and simulations
by Yuan~\cite{Yua99}, which predict a steady interface between two shear bands 
in this situation. However, we agree quantitatively with results of 
Renardy~\cite{Ren95}, who happened to look at a narrow region of high shear 
and found stability to long waves (although the paper focuses on a short-wave 
instability whose mechanism is likely to be affected by diffusion). Our 
results are in full agreement with numerical stability calculations including 
diffusion terms~\cite{Fie05}. For typical physical parameters, for which there 
is instability to perturbations of all wavelengths in the absence of 
diffusion, we have identified the scalings at which diffusion affects the 
instability. Small diffusion on a dimensionless lengthscale $l$ will 
restabilise very long and very short waves, leaving the flow unstable to 
perturbations of moderate dimensionless wavenumber $l^{1/2} < k \ll 1^{-1}$. 

We have identified two driving forces for the instability: discontinuity of 
shear rate and of normal stress across an interface. The interplay between 
these mechanisms, even for long waves, is not fully understood.

This widespread instability suggests that the existing theoretical picture 
of two stable shear bands separated by a steady interface needs further 
thought. Indeed, this result is consistent with accumulating evidence for 
erratic fluctuations~\cite{WunColLenArnRou01,BanBasSoo00,HolLopCal03,HBP98,%
FisWheFul02} in several different shear banding systems.

Future work will investigate the behaviour of the interface in the nonlinear
r\'egime, beyond the validity of this linear study. One possibility is that 
the instability saturates at a small but finite amplitude --- 
indeed, our preliminary investigations suggest that this is the case. 
This would be consistent with a narrowly localised but still unsteady 
interface, which might be interpreted as steady in experiments that did not 
have high spatial resolutions. This might even reconcile early data showing 
apparently steady interfaces with recent work revealing fluctuations. 

However, if this is not the case, then the use of a Johnson-Segalman type 
constitutive model, with or without stress diffusion, can almost never 
produce agreement with any steady banded structure observed experimentally. 
One would then need a new theoretical picture to incorporate the observed 
shear-banding effects within a stable flow which is either steady or undergoes 
only small-amplitude oscillations. 

\begin{ack}
The authors would like to thank the referees for their helpful comments.
\end{ack}

\appendix
\section{Calculation of the order $l$ contribution to $\omega$ in the 
long-wave limit}
\label{app:lowklowl}

In this section we derive a scaling form for the (negative) growth rate of 
a perturbation at $k=0$ for small $l$, in the case $\Delta=0$, \ie\ the case 
in which the stress diffusion is added through diffusion of the polymer extra 
stress term. This negative growth appears at order $l$. 

Let us return to the full governing equations 
(\ref{eq:goveq_first})--(\ref{eq:goveq_last}). We will first scale with $k$ 
and then with $l$.

Using the long-wave scalings of section~\ref{sec:longwave}, but allowing 
$\omega$ to remain order 1 (with respect to $k$), and neglecting terms of 
order $k$ yields the system
\begin{equation}
i\overline{s}_{11} + Ds_{12} = 0 \hspace{10mm}
D\overline{s}_{22} = 0 
\end{equation}
\begin{equation}
\overline{s}_{11} = \overline{s}_{22} =  -\overline{p} \hspace{10mm} 
s_{12} = \epsilon D^2\psi + t_{12} 
\end{equation}
\begin{equation}
(-i\omega + 1)t_{11} = l^2 D^2t_{11} + (1-a^2)\gd t_{12} 
+ (1-a^2)T_{12}D^2\psi
\end{equation}
\begin{equation}
(-i\omega + 1)t_{12} = l^2 D^2t_{12} - \gd t_{11} + (1-T_{11})D^2\psi
\end{equation}
As before, we solve the momentum equations to have 
\begin{equation}
s_{12} = Ay + B \hspace{10mm} 
\overline{s}_{11} = \overline{s}_{22} = -\overline{p} = iA.   
\end{equation}
We also scale the eigenvalue: $\omega=il\sigma_0 + O(l^2)$. 

We now divide the flow into three regions: the two ``outer'' regions where 
the base state shear rate and stresses are constant, and the ``inner'' region 
where base state quantities have derivatives of order $l^{-1}$. We denote 
by $X_L$ the value of a quantity in the low-shear-rate band near the wall at
$y=0$, and by $X_H$ its value in the high-shear-rate band near the wall at 
$y=1$. As in section~\ref{sec:longwave}, we will use the marginal 
viscosity 
\begin{equation}
\overline{\mu} = \epsilon + \frac{(1 - 2T_{11})}{(1+(1-a^2)\gd^2)}.
\end{equation}

In each outer region, where derivatives are order 1, we neglect terms of order 
$l$ and solve to have 
\begin{equation}
t^L_{12} = \frac{[Ay+B](\mua-\epsilon)}{\mua}
\hspace{10mm}
t^L_{11} = \frac{2[Ay+B](1-a^2)T^L_{12}}{\mua[1 + (1-a^2)\gd_L^2]}
\end{equation}
\begin{equation}
D^2\psi_L = [Ay + B]/\mua  \hspace{10mm}
\psi_L = [Ay^3 + 3By^2]/6\mua  
\end{equation}
in the low-shear band, and in the high-shear band,
\begin{equation}
t^H_{12} = \frac{[Ay+B](\mub-\epsilon)}{\mub}
\hspace{10mm}
t^H_{11} = \frac{2[Ay+B](1-a^2)T^H_{12}}{\mub[1 + (1-a^2)\gd_H^2]}
\end{equation}
\begin{equation}
D^2\psi_H = [Ay+B]/\mub \hspace{10mm} 
\psi_H = [A(y^3-3y+2)+3B(y-1)^2]/6\mub.
\end{equation}

Within the inner region, we scale lengths as $\xi = (y-\kappa)/l$ so that 
$lD = \Dxi$ and pose the series 
\begin{equation}
t_{ij} \sim l^{-1}t_{ij}^{-1} + t_{ij}^0 + \cdots \hspace{10mm} 
\psi \sim \psi^0 + l\psi^1 + l^2\psi^2 + \cdots  
\end{equation}
Matching these quantities to the outer solutions for large $|\xi|$ yields the 
following conditions: 
\begin{equation}
t_{ij}^{-1} \to 0, \mbox{ } \Dxi\psi^0 \to 0, \mbox{ } \Dxi^2\psi^1\to 0 
\mbox{ as }\xi\to\pm\infty;  
\label{eq:appmatch}
\end{equation}
\begin{equation}
t_{ij}^{0} \to t_{ij}^L(\kappa), \mbox{ } 
\psi^0 \to \psi_L(\kappa), \mbox{ } \Dxi\psi^1\to D\psi_L(\kappa), 
\mbox{ } \Dxi^2\psi^2\to D^2\psi_L(\kappa) 
\label{eq:appmatch2}
\end{equation}
as $\xi\to-\infty$; and as $\xi\to\infty$, 
\begin{equation}
t_{ij}^{0} \to t_{ij}^H(\kappa), \mbox{ } 
\psi^0 \to \psi_H(\kappa), \mbox{ } \Dxi\psi^1\to D\psi_H(\kappa), 
\mbox{ } \Dxi^2\psi^2\to D^2\psi_H(\kappa). 
\label{eq:appmatch3}
\end{equation}

We collect orders of $l$ in the resulting equations. At order $l^{-2}$ we have 
simply 
\begin{equation}
\Dxi^2\psi^0 = 0  
\end{equation}
which, along with (\ref{eq:appmatch}) gives $\psi^0 = \alpha_0$. At order 
$l^{-1}$ we have 
\begin{equation}
\epsilon \Dxi^2\psi^1 + t^{-1}_{12} = 0
\label{eq:apphom1}
\end{equation}
\begin{equation}
\Dxi^2t^{-1}_{11} - t^{-1}_{11} + (1-a^2)\gd t^{-1}_{12} 
+ (1-a^2)T_{12}\Dxi^2\psi^1 = 0
\end{equation}
\begin{equation}
\Dxi^2t^{-1}_{12} - t^{-1}_{12} - \gd t^{-1}_{11} + (1-T_{11})\Dxi^2\psi^1
= 0
\label{eq:apphom2}
\end{equation}
which are solved by 
\begin{equation}
t_{11}^{-1} = -\delta \Dxi T_{11} \hspace{10mm}  
t_{12}^{-1} = -\delta \Dxi T_{12} \hspace{10mm}  
\Dxi^2\psi^1 = -\delta \Dxi\gd.
\end{equation}
Substituting the matching conditions from 
(\ref{eq:appmatch2})--(\ref{eq:appmatch3}) and the outer solutions leads 
to the condition
\begin{equation}
A\kappa + B = 
\frac{4\delta[\mub\kappa^3- \mua(\kappa-1)^3]\mua\mub(\gd_H-\gd_L)}
{[\mub\kappa^2 - \mua(\kappa-1)^2]^2 - 4\mua\mub\kappa(\kappa-1)}.
\end{equation}

Finally, at order 1 our remaining equations are 
\begin{equation}
\epsilon \Dxi^2\psi^2 + t^0_{12} = A\kappa +B
\end{equation}
\begin{equation}
\Dxi^2t^0_{11} - t^0_{11} + (1-a^2)\gd t^0_{12} 
+ (1-a^2)T_{12}\Dxi^2\psi^2 = 
- \sigma_0\delta \Dxi T_{11} 
\end{equation}
\begin{equation}
\Dxi^2t^0_{12} - t^0_{12} - \gd t^0_{11} + (1-T_{11})\Dxi^2\psi^2
= - \sigma_0\delta \Dxi T_{12} 
\end{equation}
Where equations (\ref{eq:apphom1})--(\ref{eq:apphom2}) were homogeneous, these
are inhomogeneous ODEs with a forcing on the RHS which comes from the previous
order calculation. Together with the conditions on $t_{ij}^0$ from 
(\ref{eq:appmatch2}) and (\ref{eq:appmatch3}), they provide a constraint on 
$\sigma_0$. 

Although we cannot solve this problem, we can reduce its complexity. If we 
define 
\begin{equation}
\alpha = \Dxi^2\psi^2/[A\kappa+B] \hspace{10mm} 
\tilde{t}_{ij} = t^0_{ij}/[A\kappa + B] \hspace{10mm} 
\tilde{\sigma} = -\sigma_0\delta/[A\kappa + B]   
\end{equation}
then the governing equations become 
\begin{equation}
\epsilon\alpha + \tilde{t}_{12} = 1
\end{equation}
\begin{equation}
\Dxi^2\tilde{t}_{11} - \tilde{t}_{11} + (1-a^2)\gd \tilde{t}_{12} 
+ (1-a^2)T_{12}\alpha = \tilde{\sigma}\Dxi T_{11}
\end{equation}
\begin{equation}
\Dxi^2\tilde{t}_{12} - \tilde{t}_{12} - \gd \tilde{t}_{11} 
+ (1-T_{11})\alpha = \tilde{\sigma}\Dxi T_{12}
\end{equation}
and the matching conditions,
\begin{equation}
\tilde{t}_{11} \to 
\frac{2(1-a^2)T^L_{12}}{\mua[1 + (1-a^2)\gd_L^2]} \hspace{10mm}  
\tilde{t}_{12} \to \frac{\mua-\epsilon}{\mua}
\mbox{ as }\xi\to-\infty
\end{equation}
\begin{equation}
\tilde{t}_{11} \to 
\frac{2(1-a^2)T^H_{12}}{\mub[1 + (1-a^2)\gd_H^2]} \hspace{10mm}  
\tilde{t}_{12} \to \frac{\mub-\epsilon}{\mub}
\mbox{ as }\xi\to\infty.
\end{equation}
Neither the governing equations nor the matching conditions have any dependence
on $\kappa$, the position of the interface between the two shear rate 
bands. Thus there is a single value of $\tilde{\sigma}$ for each set of 
parameters $\{a,\epsilon,T_b\}$ independent of $U_\mathrm{wall}$ and $\kappa$.
The long-wave term of the eigenvalue $\omega$ is then given by 
\begin{equation}
\omega \sim 
\frac{-4il[\mub\kappa^3- \mua(\kappa-1)^3]\mua\mub(\gd_H-\gd_L)\tilde{\sigma}}
{[\mub\kappa^2 - \mua(\kappa-1)^2]^2 - 4\mua\mub\kappa(\kappa-1)}
+ O(l^2) + O(k). 
\end{equation}

For the results given in figure 3 of \cite{Fie05}, we had $a=0.3$, 
$\epsilon=0.05$ and $T_b=0.506158$, giving shear rates of $\gd_L = 0.66143$ 
and $\gd_H=7.0893$. The interface position was $0.79176$ and a fit of the data
gave $\omega \sim -10l$. We can deduce that 
\begin{equation}
\tilde{\sigma}(a=0.03,\epsilon=0.05,T_b=0.506158) \approx 14. 
\end{equation}

\bibliography{papers}
\bibliographystyle{elsart-num}

\end{document}